\documentclass[12pt]{article}

\usepackage{color}		% Non-standard graphics support
\usepackage{graphicx}	% Non-standard graphics support
\usepackage{amsbsy}		% Bolds (vectors, tensors, etc.)
\usepackage{amsfonts}		% AMS fonts
\usepackage{amsmath}		% AMS maths
\usepackage{amssymb}		% AMS symbols
\usepackage{wasysym}
\usepackage{multirow}
\usepackage{mciteplus}
\usepackage{psfrag}
\usepackage{url}

\newcommand{\sinb}{\mathrm{s}_\beta}
\newcommand{\cosb}{\mathrm{c}_\beta}

\newcommand*{\Ave}[1]{\mathinner{\left\langle{#1}\right\rangle}}
\newcommand{\GeV}{\;\mathrm{GeV}}
\newcommand{\TeV}{\;\mathrm{TeV}}
\newcommand{\mue}{\mu_{\mathrm{eff}}}

\newcommand{\NMT}{{\tt NMSSMTools}}
\newcommand{\HB}{{\tt HiggsBounds}}
\newcommand{\SI}{{\tt SuperIso}}

\newcommand{\CP}{$\mathcal{CP}$}

\begin{document}
\hfill {\tt CERN-PH-TH/2010-304}

\hfill {\tt DESY 10-246}

\hfill {\tt LU-TP 10-29}

\def\thefootnote{\fnsymbol{footnote}}

\begin{center}
\Large\bf\boldmath
\vspace*{0.8cm} Light Higgs bosons in phenomenological NMSSM
\unboldmath
\end{center}
\vspace{0.6cm}
\begin{center}
F.~Mahmoudi$^{1,2,}$\footnote{Electronic address: mahmoudi@in2p3.fr},
J.~Rathsman$^{3,4,}$\footnote{Electronic address: johan.rathsman@fysast.uu.se}, 
O.~St{\aa}l$^{5,}$\footnote{Electronic address: oscar.stal@desy.de},
L.~Zeune$^{5,6,}$\footnote{Electronic address: lisa.zeune@desy.de} \\[0.4cm] 
\vspace{0.6cm}
{\sl $^1$ CERN Theory Division, Physics Department\\ CH-1211 Geneva 23, Switzerland}\\[0.2cm]
{\sl $^2$ Clermont Universit{\'e}, Universit\'e Blaise Pascal, CNRS/IN2P3,\\
LPC, BP 10448, 63000 Clermont-Ferrand, France}\\[0.2cm]
{\sl $^3$ High-Energy Physics, Department of Physics and Astronomy\\ Uppsala University, Box 535, S-751\,21 Uppsala, Sweden}\\[0.2cm]
{\sl $^4$ Theoretical High Energy Physics\\
 Department of Astronomy and Theoretical Physics\\
Lund University, S{\"o}lvegatan 14A, S-223\,62 Lund, Sweden}\\[0.2cm]
{\sl $^5$ Deutsches Elektronen-Synchrotron DESY\\
 Notkestra{\ss}e 85, D-22607 Hamburg, Germany}\\[0.2cm]
{\sl $^6$ Universit{\"a}t G{\"o}ttingen, II.~Physikalisches Institut\\
D-37077 G{\"o}ttingen, Germany }
\end{center}

\renewcommand{\thefootnote}{\arabic{footnote}}
\setcounter{footnote}{0}

\begin{abstract}
We consider scenarios in the next-to-minimal supersymmetric model (NMSSM) where the CP-odd and charged Higgs bosons are very light. As we demonstrate, these can be obtained as simple deformations
of existing phenomenological MSSM benchmarks scenarios with parameters defined at the weak scale. This  offers a direct and meaningful comparison to the MSSM case. Applying a wide set of up-to-date constraints from both high-energy collider and flavour physics, the Higgs boson masses and couplings are studied in viable parts of parameter space. The LHC phenomenology of the light Higgs scenario for neutral and charged Higgs boson searches is discussed.%
\end{abstract}
\newpage

\section{Introduction}
The explanation behind electroweak symmetry breaking (EWSB) is still largely unknown. Assuming the symmetry is spontaneously broken through the Higgs mechanism \cite{Higgs:1964ia,*Englert:1964et,*Higgs:1964pj} still leaves room for speculation on the structure of the Higgs sector. Extending the standard model (SM) with softly broken supersymmetry at the TeV scale has several benefits: it protects the scalar masses from being sensitive to the large hierarchy of scales through quadratic divergences, it allows for gauge coupling unification, and it presents a candidate for the cold dark matter of the universe in terms of a stable lightest supersymmetric particle (LSP). 

In the minimal supersymmetric standard model (MSSM), the electroweak symmetry is broken by the presence of two complex Higgs doublets. The MSSM Higgs sector has a rich phenomenology which has been explored in exquisite detail \cite{Djouadi:2005gj}. Searches for MSSM Higgs bosons constitute a cornerstone of experimental particle physics at colliders. Driven in parallel by the negative results from direct searches, and indirect results from the measurement of e.g.~electroweak precision observables, limits can be placed on their masses and couplings. The combined data implies that a SM-like Higgs boson should be just around the corner \cite{LEPEWWG}, and definitely within the reach of the LHC. There are also theoretical arguments favouring a lightest Higgs mass $m_h\lesssim 130~\mathrm{GeV}$ in the MSSM, where $m_h<m_Z$ hold at tree-level. It has even been argued \cite{Ellis:2007fu}, that the MSSM gives a better combined fit to the
available data than does the SM, adding further support to the idea of weak scale supersymmetry.

Despite the great interest in the minimal model, we find it advisable to keep an
open mind also with regards to further enlargements of the Higgs sector. After all,
there exists no direct evidence for any particular theory beyond the standard model. It must therefore be ensured that no experimental opportunities are lost due to ignorance of
parameter regions which are excluded in the MSSM. Corresponding parameter regions may be allowed in other models, as we shall demonstrate explicitly in this work for the simplest extension of the MSSM, the next-to-minimal supersymmetric standard model (NMSSM). For a general introduction to this model, we recommend the recent reviews \cite{Maniatis:2009re,Ellwanger:2009dp}. Details necessary for this paper, in particular on the extended Higgs sector, will be described below.

Let us briefly recall the main theoretical motivation for the NMSSM. The MSSM superpotential $W^{(d)}$ contains, in addition to the Yukawa terms of mass dimension $d=3$, the
bilinear coupling $W^{(2)}_{\mathrm{MSSM}}=\mu \hat{H}_u\cdot \hat{H}_d$
of the two Higgs doublet superfields.\footnote{We use the notation where $\hat{\Phi}$ denotes a chiral superfield and $\Phi$ its scalar component. The dot gives the $SU(2)$-invariant scalar product such that $\hat{H}_u\cdot \hat{H}_d=H_u^+H_d^--H_u^0H_d^0$.} The parameter $\mu$ has positive mass dimension, but since it is introduced directly in the supersymmetric theory it is not associated with any obvious scale (e.g.~the SUSY breaking scale). This leaves only $\mu=0$ or $M_{\mathrm{GUT}}$ as natural choices,  both of which are phenomenologically impossible. An acceptable phenomenology requires $\mu$ close to the scale of electroweak symmetry breaking: $\mu\sim~\mu_{\mathrm{EW}}< M_{\mathrm{SUSY}}$, where $M_{\mathrm{SUSY}}$ is the scale of supersymmetry breaking. This is known as the `$\mu$ problem' of the MSSM. The primary motivation for the NMSSM is to solve this problem. Replacing $W^{(2)}_\mathrm{MSSM}$ with a coupling of the Higgs doublets to a new singlet field $\hat{S}$, the $\mu$ parameter is generated dynamically from soft terms as the vacuum expectation value of $S$. The NMSSM can also help to reduce the required amount of fine-tuning compared to the MSSM \cite{Dermisek:2009si}. 

The NMSSM does not only have favourable properties from a theoretical perspective, but it also has an interesting Higgs phenomenology which  extends that of the MSSM. The singlet components mix with the Higgs scalars (and neutralinos), modifying the tree-level mass relations, and introducing new modes of Higgs production and decay which can be relevant for the LHC. The present work starts from the MSSM limit of the NMSSM, which will be discussed further in the next section. More specifically, we consider low-energy MSSM benchmark scenarios as our starting point. These are easy to work with due to the relatively low number of free parameters; in the Higgs sector typically the CP-odd Higgs mass $m_A$ (or equivalently $m_{H^\pm}$) and $\tan\beta$. The simplicity inherent in the benchmarking approach is appreciated by experimental collaborations, sometimes to the degree that one could fear loss of information in the process of reporting results. One of the aims of this paper is therefore to use present experimental limits to determine how our current understanding of well-known scenarios in the MSSM is modified when going to the NMSSM. To this end, we use input from LEP, the Tevatron, and flavour physics experiments performed at lower energies to constrain the parameter space. Of special interest to us is the possibility to have comparably light Higgs bosons, while still being compatible with all known limits. This question is investigated in more detail, focusing in particular on the cases where a CP-odd Higgs and/or the charged Higgs boson is very light. For scenarios that fulfil the experimental constraints, we calculate and discuss Higgs decay modes and production cross sections for LHC running at $14$~TeV.

This work does not concern the idea of a hidden NMSSM Higgs sector, or how to establish a 'no-loose' theorem under these circumstances  \cite{Ellwanger:2001iw,*Ellwanger:2003jt,*Ellwanger:2004gz,*Moretti:2006hq,*Moretti:2006sv,*Forshaw:2007ra,Belyaev:2008gj}. Quite to the contrary, we focus on scenarios with one or more light Higgs bosons, maximizing the accessible phenomenology while avoiding exclusion by the present data. Work along similar lines is presented in \cite{Dermisek:2008uu}, with the motivation of explaining the slight excess of Higgs-like events at LEP observed around $m_h\simeq 90$~GeV as $H_1$ production followed by the dominant (but unobserved) decay $H_1\to A_1A_1$ \cite{Dermisek:2005ar,*Dermisek:2005gg,*Dermisek:2006wr,*Dermisek:2007yt,Dermisek:2010mg}. The phenomenology of these so-called ideal Higgs scenarios---although obtained through a different strategy---should be closely related to some of the scenarios we investigate here. 
 In addition to our different starting point, we extend the earlier results by including new limits from Higgs searches at the Tevatron in our analysis, as well as the charged Higgs boson searches at LEP and the Tevatron. We also present combined results from collider and flavour constraints. Our independent implementation of the latter serves as an important cross-check and an update on the Higgs constraints in the NMSSM. Finally, we show explicitly the effects of these constraints on the quantities of phenomenological interest at colliders. Unlike the NMSSM benchmark scenarios presented in \cite{Djouadi:2008uw}---where the charged Higgs boson is always taken to be heavy---we also consider the case when it is light; an interesting possibility discussed in \cite{Akeroyd:2007yj}.

This paper is organized as follows: in Section~\ref{sect:Higgs} we introduce the NMSSM Higgs sector and describe briefly the computation of the Higgs masses and mixing at tree-level. This is followed in Section~\ref{sect:Analysis} by a more detailed description of our analysis strategy. Section~\ref{sect:Const} first presents the experimental input, which is then applied to give up-to-date exclusion limits on the NMSSM Higgs boson masses. The LHC phenomenology for the Higgs sector of models allowed by the combined constraints is discussed in Section \ref{sect:LHC}. Finally, Section \ref{sect:Summary} contains a summary and the conclusions.

\section{The Higgs sector of the NMSSM}
\label{sect:Higgs}
To simplify the analysis, a common restriction of the most general supersymmetric model with a gauge singlet is to consider a scale invariant superpotential. This model is sometimes referred to as the $Z_3$--symmetric NMSSM. With this assumption, the superpotential takes the form
\begin{equation}
W_{\mathrm{NMSSM}}=W^{(3)}_{\mathrm{MSSM}}+\lambda \hat{S}\hat{H}_u\cdot
\hat{H}_d+\frac{1}{3}\kappa \hat{S}^3,
\label{eq:W_NMSSM}
\end{equation}
with all parameters dimensionless. $W^{(3)}_{\mathrm{MSSM}}$ contains the Yukawa terms of the MSSM which are not modified. The supersymmetric parts of the Higgs potential are derived by evaluating the usual $F$-- and $D$--terms. In addition to the superpotential, a complete phenomenological model requires specification of the soft SUSY-breaking potential to have $V_{\mathrm{Higgs}}=V_F+V_D+V_{\mathrm{soft}}$. The terms in $V_{\mathrm{soft}}$ containing only the Higgs fields are given by
\begin{eqnarray}
V_{\mathrm{soft}}&=&m_{H_u}^2|H_u|^2+m_{H_d}^2|H_d|^2+m_{S}^2|S|^2\nonumber\\
&& + \bigl(\lambda
A_\lambda SH_u\cdot H_d +\frac{1}{3}\kappa A_\kappa S^3+\mathrm{h.c.}\bigr).
\label{eq:soft}
\end{eqnarray}
The remaining terms in $V_\mathrm{soft}$ keep their MSSM expressions.
In addition to the unstable doublet minimum required for EWSB, it is assumed that $m_S^2\lesssim A_\kappa^2/9$ (which may be radiatively generated) to make $S=0$ unstable. Expanding the singlet field $S=v_s+\frac{1}{\sqrt{2}}\left(\phi_s+\mathrm{i}\sigma_s\right)$ around its vacuum expectation value
$v_s\equiv\Ave{S}$ gives rise to an effective parameter $\mue=\lambda v_s$ with mass dimension. Since $v_s$ originates from SUSY breaking, a phenomenologically viable value $|\mu_{\mathrm{eff}}|\lesssim M_{\mathrm{SUSY}}$ is natural.
Having noted this important dynamical mechanism to generate the $\mu$ parameter, we drop the `eff' subscript in the following.

Following EWSB, the physical spectrum contains seven Higgs scalars, five of which are neutral, and a single charged pair. If \CP\ is conserved\footnote{Unlike the MSSM, automatic \CP\ conservation at tree-level is not guaranteed, but constitutes an additional assumption. In the \CP\ conserving case it is generically possible to use real values for all parameters appearing in the Higgs potential.} (as assumed here) the neutral mass eigenstates are arranged into three CP-even Higgs bosons $(H_1,H_2,H_3)$, ordered in mass such that $m_{H_1}\leq m_{H_2} \leq m_{H_3}$, and two CP-odd states $(A_1,A_2)$ with $m_{A_1}\leq m_{A_2}$. The charged Higgs boson---which makes a proper mass eigenstate also in the presence of \CP\ violation---is denoted by $H^\pm$. Using the minimization conditions for the potential, the scalar mass
parameters of Equation~\eqref{eq:soft} can be eliminated in favour of the vacuum expectation values of the doublets. We define as usual
\begin{equation*}
v_u^2+v_d^2=v^2\simeq (174\GeV)^2,\qquad \tan\beta=\frac{v_u}{v_d}.
\end{equation*}
Due to the additional interactions with the NMSSM singlet that have been introduced, a full specification of the
Higgs sector now requires six parameters at tree-level:
\begin{equation*}
\lambda \quad \kappa \quad A_\lambda \quad A_\kappa \quad \tan\beta  \quad v_s.
\end{equation*}
Conventions can be chosen such that $\lambda$ and $\tan\beta$ are positive, while the remaining parameters can have either sign.

The presentation of the Higgs mass matrices at tree-level is simplified by defining an effective doublet mass $m_A$ as
\begin{equation}
m_A^2\equiv \frac{\lambda v_s}{\sinb\cosb}\Bigl(A_\lambda+\kappa v_s\Bigr),
\label{eq:mA2}
\end{equation}
with $\cosb=\cos\beta$, and $\sinb=\sin\beta$. Note that $m_A$ in general does not correspond to the mass of any physical Higgs boson in the NMSSM. Eliminating the massless degree of freedom eaten by the $Z$, the remaining $2\times 2$ mass matrix for the pseudoscalars can be written in the basis $(A, \sigma_s)$, where $A= \mathrm{Im}(\cos\beta H_u+ \sin\beta H_d)$  is the doublet component. In this basis, it is given by
\begin{eqnarray}
&&\mathcal{M}_P^2=\nonumber\\
&&\hspace*{-0.9cm}\left(\begin{array}{cc}
\displaystyle m_A^2 & \displaystyle \frac{v}{v_s}\left(m_A^2\cosb\sinb-3\lambda\kappa v_s^2\right) \\
\displaystyle \frac{v}{v_s}\left(m_A^2\cosb\sinb-3\lambda\kappa v_s^2\right) & \displaystyle
\frac{v^2\cosb\sinb}{v_s^2}\left(m_A^2\cosb\sinb+3\lambda\kappa v_s^2\right)-3\kappa A_\kappa v_s
\end{array}
\right)\nonumber
\end{eqnarray}
The pseudoscalar masses are diagonalized by a simple rotation with an angle $\theta_A$ such that
\begin{equation}
\left\{\begin{array}{l}
A_1=A\cos\theta_A+\sigma_s\sin\theta_A \\
A_2=-A\sin\theta_A+\sigma_s\cos\theta_A,
\end{array}
\right. 
\end{equation}
where an explicit relation for the mixing angle is
\begin{equation}
\cos\theta_A=\frac{\mathcal{M}_{P,12}^2}{\sqrt{\mathcal{M}_{P,12}^4+\left(m_{A_1
}^2-\mathcal{M}_{P,11}^2\right)^2}}.
\end{equation}
To describe the mixing in the CP-even Higgs sector requires the introduction of a full $3\times 3$ unitary matrix. In a basis $S^{\rm weak}=(\mathrm{Re}(H_u),\mathrm{Re}(H_d),\mathrm{Re}(S))$ where the mass eigenstates $H_i$ are given by $H_i=S_{ij}S^{\rm weak}_j$ we use the parametrisation
\begin{equation}
S=\begin{pmatrix} c_{12}c_{13} & s_{12}c_{13} & s_{13} \\
                  -s_{12}c_{23}-c_{12}s_{13}s_{23} & c_{12}c_{23}-s_{12}s_{13}s_{23} & c_{13}s_{23} \\
                  -c_{12}c_{23}s_{13}+s_{12}s_{23} & -c_{12}s_{23}-s_{12}s_{13}c_{23} & c_{13}c_{23} 
\end{pmatrix},                  
\end{equation}
with $c_{ij}=\cos\theta_{ij}$ and $s_{ij}=\sin\theta_{ij}$.\footnote{This corresponds to a mixing in the MSSM limit which takes the form $\theta_{12}\rightarrow -\alpha$, $\cos\theta_{13}\to 1$, and $\cos\theta_{23}\to 1$.} For brevity, we do not give the entries of the CP-even mass matrix $\mathcal{M}_S$ here. They can be found for example in \cite{Ellwanger:2009dp}. The tree-level value of $m_{H_1}$ is no longer limited from above by $m_h^2\leq m^2_Z\cos^2 2\beta$ (as is the case in the MSSM). Instead the modified limit \cite{Drees:1988fc}
\begin{equation}
m_{H_1}^2\leq m_Z^2\cos^2 2\beta +\frac{\lambda^2m_W^2}{g^2}\sin^2 2\beta
\end{equation} 
applies. Finally, the tree-level mass of the physical charged Higgs boson is given by
\begin{equation}
m_{H^\pm}^2=m_A^2+m_W^2-\lambda^2v^2,
\label{eq:mHc}
\end{equation}
where $m_A^2$ is given by Equation~\eqref{eq:mA2}. For a fixed doublet mass, the term $-\lambda^2v^2$ 
reduces the charged Higgs mass compared to its MSSM value. 

The whole discussion so far concerned the Higgs masses at tree-level. Like in the MSSM, the tree-level relations may be subject to sizeable higher order corrections \cite{Degrassi:2009yq,Staub:2010ty}. The Higgs mass corrections taken into account in our numerical analysis are discussed further in the next section. As a final point, let us discuss the MSSM limit. This is obtained by taking simultaneously $\lambda\to 0$ and $\kappa\to 0$, keeping the ratio $\kappa/\lambda$ constant, thus decoupling the singlet from interacting with the Higgs doublets. A finite $|\mu|>0$ in this limit (as required to get massive charginos) is achieved by taking $v_s\to \infty$. The remaining dimensionful parameters are held constant.
This illustrates the possibility to consider a class of models with a continuous transition from the MSSM to the NMSSM.
Finally, we note the interesting fact that even if all influence on Higgs physics vanishes the LSP may be singlino-like, giving a possibly long-lived (or even charged) NLSP which could lead to modified collider phenomenology even in the decoupling limit \cite{Ellwanger:1997jj}.

\section{Analysis Strategy}
\label{sect:Analysis}
The masses (and mixing matrices) of the Higgs bosons receive numerically important corrections at higher orders. These loop corrections introduce a dependence in the Higgs sector on the soft SUSY-breaking parameters of other sectors, most notably from stops (for high $\tan\beta$ also sbottoms), which have sizeable
couplings to the Higgs doublets. Both the dominant corrections proportional to the Yukawa couplings $y_t^4$, $y_b^4$ and the mixed electroweak/Yukawa terms of order $g^2y_{t}^2$, $g^2y_b^2$ are similar between the MSSM and the NMSSM. For the numerical analysis of the NMSSM Higgs sector we use the code {\tt NMSSMTools 2.2.0} \cite{Ellwanger:2005dv,*Ellwanger:2004xm}, which includes the dominant one-loop and leading logarithmic two-loop corrections to the Higgs masses.

To study how constraints on the Higgs sector are modified when going from the MSSM to the NMSSM, we adopt a strategy to fix the higher order corrections from sparticles in order to focus on `genuine' NMSSM effects which appear already at tree-level. To achieve this, we make use of MSSM benchmark scenarios for Higgs searches which were originally presented for LEP \cite{Carena:1999xa}, and later extended to hadron colliders \cite{Carena:2002qg}. The benchmark scenarios provide a starting point in terms of the soft SUSY-breaking parameters at the
electroweak scale. Recall that the parameters specified in these scenarios are the universal scalar mass $M_\mathrm{SUSY}$, the gaugino masses $M_2$ and $M_3$, with $M_1$ typically given by the GUT relation $M_1=\frac{5s^2_W}{3c^2_W}M_2$, the trilinear couplings for the third generation $A_t=A_b=A_\tau$, and the value for $\mu$. Once everything is fixed in this way, only $m_A$ (or $m_{H^\pm}$) and $\tan\beta$ remain as free parameters. 

\begin{table}
\centering
\begin{tabular}{ccc}
\hline
\hspace{10pt}Parameter\hspace{10pt} & \hspace{10pt}Scenario (A)\hspace{10pt} &
\hspace{10pt}Scenario (B)\hspace{10pt}   \\
\hline
$M_{\mathrm{SUSY}}$ & $1\;\mathrm{TeV}$ & $1\;\mathrm{TeV}$ \\
$X_t^{\overline{\mathrm{MS}}} $ & $-\sqrt{6}\,M_{\mathrm{SUSY}}$ & $0$\\
$\mu$ & $200\GeV$ & $200\GeV$ \\
$M_2$ & $200\GeV$ & $200\GeV$ \\
$M_3$ & $800\GeV$ & $800\GeV$ \\
\hline
\end{tabular}
\caption{Values for the soft SUSY-breaking parameters and $\mu$ in the MSSM
benchmark scenarios (A) \emph{constrained maximal mixing} and (B) \emph{no mixing} of
\cite{Carena:2002qg}. In the on-shell renormalization scheme for $m_A$,
$|X_t^{\mathrm{OS}}|=2\,M_{\mathrm{SUSY}}$ for maximal mixing.}
\label{tab:scenarios}
\end{table}

Since we are interested in Higgs mass limits, we look at benchmark scenarios covering the two extremes of radiative corrections to the lightest Higgs mass: one scenario with maximal mixing in the stop sector and one with no mixing. The full parameter sets for the two scenarios we consider are given in Table~\ref{tab:scenarios}. Scenario (A) is the \emph{constrained maximal mixing} scenario and scenario (B) is called \emph{no mixing}. Mixing refers to the size of the off-diagonal entries in the sfermion mass matrices, which for the stops are given by $m_tX_t$ where $X_t=A_t-\mu\cot\beta$. When the mixing is maximal, it also maximizes the one-loop corrections to the lightest Higgs mass. Including also two-loop corrections, the value of $X_t$ for
which this is achieved depends on the renormalization scheme; see \cite{Carena:2000dp} for a
discussion and comparison between $\overline{\mathrm{MS}}$ and OS (on-shell)
results. It has been shown that certain two-loop terms lift the degeneracy of the $m_h$ maximum under $X_t\to -X_t$. This is of some significance to us here, since we consider the
\emph{constrained} maximal mixing scenario which differs from maximal mixing
only in the sign of $X_t$. Even though this gives a lower maximum value for $m_{H_1}$
(by up to $5\GeV$), this scenario is in better agreement with data from $b\to
s\gamma$ transitions.\footnote{The main sparticle contribution to
$\mathrm{BR}(B\to X_s\gamma)$ is proportional to $\mathrm{sgn}(\mu A_t)$. Since
the other major contribution (from $H^\pm$) is always positive, the negative
sign for $A_t$ is required for their successful cancellation.} A negative  $X_t$ at the weak scale is also strongly favoured in constrained MSSM scenarios with parameters evolved from the scale of grand unification. For the scenario with no mixing, $X_t=X_b=X_\tau=0$ is imposed. To calculate the MSSM Higgs spectrum we use the code {\tt FeynHiggs 2.7.4} \cite{Frank:2006yh,*Degrassi:2002fi,*Heinemeyer:1998np,*Heinemeyer:1998yj}.

Our strategy for the NMSSM case is to extend the MSSM benchmarking approach in the following manner: we keep the soft parameters fixed at
the values given by the MSSM scenario. Also the effective value of
$\mu=\lambda v_s$ is kept at its MSSM value. As before, this leaves the tree-level Higgs sector free. Since the CP-odd Higgs mass parameter $m_A$ is not physical in the NMSSM, we would like to use $m_{H^\pm}$ as input. This is achieved by using Equations~\eqref{eq:mA2} and \eqref{eq:mHc} to determine $A_\lambda$ at tree-level, which is then corrected at higher orders (using {\tt NMSSMTools}) through iteration.
In summary, the NMSSM scenarios have three additional parameters compared to the MSSM:
\begin{equation}
(m_{H^\pm},\tan\beta)\rightarrow (m_{H^\pm},\tan\beta,\lambda,\kappa,A_\kappa).
\end{equation}
To present constraints on the NMSSM Higgs sector in a comprehensible way, we will perform scans over the additional parameters. Their domain of interest is limited from the properties of the theory. Following \cite{Miller:2003ay}, we require that $\lambda$, $\kappa$, and the Yukawa couplings (most importantly $y_t$, or for large $\tan\beta$ also $y_b$) remain perturbative to the GUT scale. This leads to the upper limit
\begin{equation}
\sqrt{\lambda^2+\kappa^2}\lesssim 0.7.
\label{eq:kappalambda}
\end{equation} 

\begin{figure}
\centering
\includegraphics[width=0.4\columnwidth]{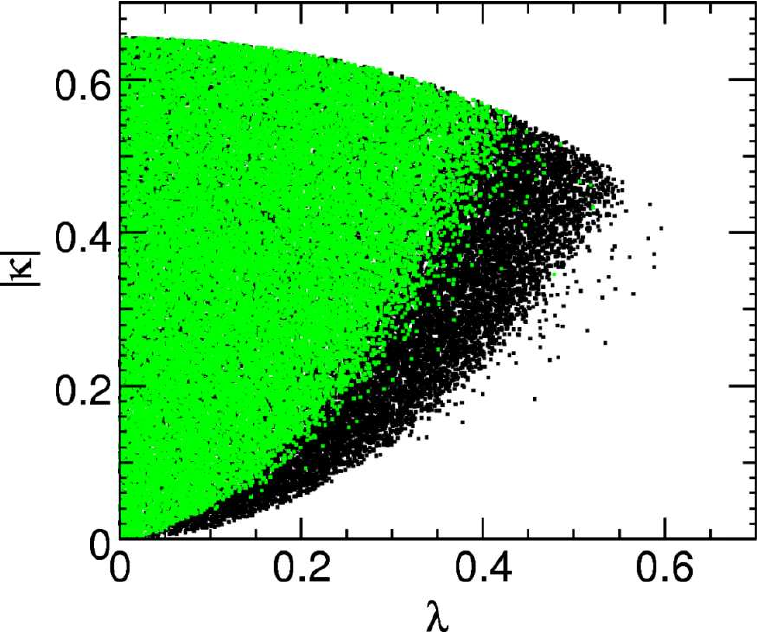}
\caption{Distribution of allowed points in the space of NMSSM couplings for scenario (A). Dark (black) points are compatible with perturbative unification, while the brighter (green) points in addition fulfil collider constraints on Higgs masses.}
\label{fig:kappalambda}
\end{figure}

This condition is illustrated in Figure~\ref{fig:kappalambda}, which shows for scenario (A) a sample of weak-scale values for $(\lambda,\kappa)$ compatible with perturbativity. The same analysis gives a lower limit on $\tan\beta\gtrsim 1.4$ using the top mass $m_t=173.3\;\mathrm{GeV}$ \cite{topmass}. We employ this value for the top quark pole mass throughout this work.

The range for $A_\kappa$ can be restricted by demanding the symmetry breaking minimum of the potential to be stable. From this can be derived \cite{Miller:2003ay} an approximate condition valid for high $\tan\beta$ and high $m_A$
\begin{equation}
-\frac{4\kappa\mu}{\lambda} \lesssim A_\kappa \lesssim 0.
\end{equation}

\begin{figure}
\centering
\includegraphics[width=0.45\columnwidth]{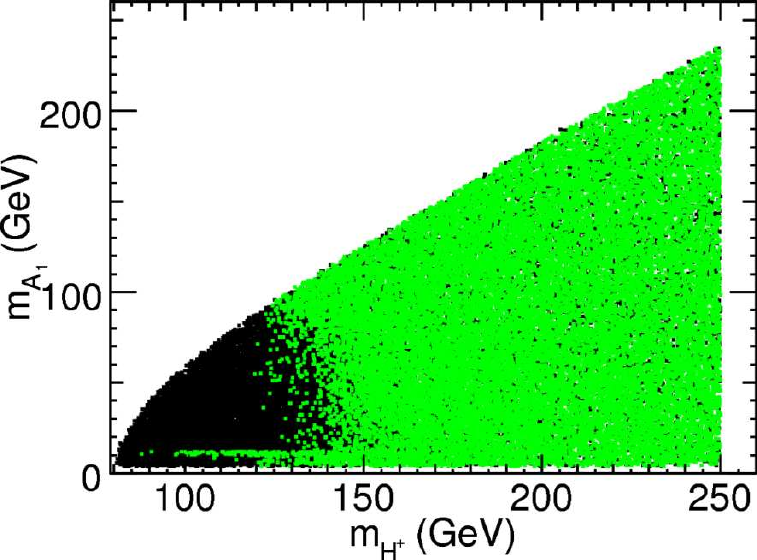}
\caption{Masses of the lightest CP-odd Higgs boson $A_1$ and charged Higgs boson in the NMSSM for scenario (A). All accepted points in the displayed mass range are shown in black, while points that are in addition compatible with collider constraints (see Section~\ref{sect:Const}) are shown in green.}
\label{fig:ma1mhc}
\end{figure}

Since this condition is only approximate, and since positive scalar mass squares are necessary for stability, we do not employ scanning over $A_\kappa$ directly. Instead we use the lightest CP-odd Higgs mass $m_{A_1}$ as input parameter and determine $A_\kappa$ through an iterative procedure. The stability of the EWSB minimum is checked numerically. Since the charged Higgs mass does not depend on $A_\kappa$ to the computed order, we can use both these masses as independent input parameters in a pseudo on-shell approach. We find that this greatly improves scanning efficiency for low Higgs masses, with about $60\%$ of the tried points giving a physically viable spectra for the scanning ranges described below. The distribution of points in the ($m_{A_1}, m_{H^\pm}$) plane obtained for scenario (A) using this method is shown in Figure~\ref{fig:ma1mhc}.

It is important to exercise caution when interpreting results from scans over multi-dimensional parameter space, especially when the sampled parameters are related non-linearly to the fundamental parameters of the theory as is the case here. In particular one should not interpret the posterior density of points in a particular region as a probability in the frequentist sense. What can be shown by the procedure employed here is the existence of valid models or---trusting the sampling is sufficiently covering---their non-existence in absence of valid points for a certain parameter region. With this objective in mind, we perform random scans over the parameter space of each benchmark scenario with uniform sampling in the ranges $m_{H^\pm}\in [80, 500]\;\mathrm{GeV}$, $\tan\beta\in [1.2,50]$, $\lambda\in [0,0.6]$, $\kappa\in [-0.7,0.7]$ (requiring Equation~\eqref{eq:kappalambda} to hold), and finally $m_{A_1}\in [4, 500]\;\mathrm{GeV}$ with the upper limit $m_{A_1}<m_A$ imposed by the tree-level relation given in Equation~\eqref{eq:mHc}. We limit ourselves to the case $m_{A_1}>2 m_{\tau}$ (where hadronic decay modes other than $b\bar{b}$ are not relevant). Constraints on an even lighter $A_1$ are presented in \cite{Andreas:2010ms}. For each scenario, we generate $\mathcal{O}(10^5)$ physical parameter points.

\section{Experimental Constraints}
\label{sect:Const}
To investigate the viability of the considered scenarios, we take into account existing constraints from several different sources as detailed below. As a general constraint, the direct limits on sparticle (not Higgs) masses compiled by the PDG \cite{Amsler:2008zzb} are checked. In the benchmark scenarios considered here, they are always fulfilled. Furthermore, to provide a possible explanation for the dark matter density of the universe, it should be required that the lightest supersymmetric particle (LSP) is neutral. Here this is the lightest neutralino which has a typical mass $m_{\chi^0_1}\sim 90\GeV$. Since the density $\Omega^{\mathrm{DM}} h^2$ of cold dark matter (CDM) in the universe is very precisely determined by WMAP \cite{Komatsu:2010fb} in the so-called concordance model, strong constraints can be derived on the SUSY parameter space under standard thermal relic assumptions for the nature of the DM. This constraint can be used to reduce the effective dimensionality of parameter space by one unit. However, we do not apply the CDM constraints here, since we are interested in the Higgs sector, and these constraints could equally well be fulfilled by making a change to the underlying MSSM parameters (similarly to \cite{Ellis:2007ka}). Moreover, it is always possible to accommodate the WMAP constraints by changing the cosmological assumptions in the early Universe (yet compatible with the cosmological observations) \cite{Arbey:2008kv,*Arbey:2009gt}. A detailed study of this topic goes beyond the scope of this paper.

\subsection{Collider constraints}
\subsubsection{LEP} 
The four LEP experiments have placed limits on the masses and couplings of neutral MSSM Higgs bosons in a large number of channels \cite{Schael:2006cr}, most of which are also relevant in the NMSSM case. Primarily two modes of neutral Higgs production were probed at LEP: 
the Higgs-strahlung process $e^+e^-\to ZH_i$, by which a single Higgs boson is produced for each event, and the associated production $e^+e^-\to H_iA_j$. The Higgs decay modes considered in the LEP analyses involve most final state combinations of $n_b$ bottom quarks and $n-n_b$ tau leptons for $n,n_b\leq 6$. There also exists an important limit on the lightest CP-even Higgs mass $m_{h}$ in the MSSM from a decay mode independent search where the mass spectrum of the recoiling $Z$ boson was analysed. We take into account the exclusion by all these channels (and many others, in total 541 channels for NMSSM case) consistently at $95\%$ CL using the code {\tt HiggsBounds 2.0.0} \cite{Bechtle:2008jh} which contains a collection of data on Higgs searches.\footnote{The LEP limits are also available in \NMT, and we have verified that applying these give results consistent with those from \HB\ within expected uncertainties.} One particular channel---not part of the original LEP analysis and therefore not yet present in \HB---is Higgs-strahlung followed by the decay $H_i\to A_1A_1\to \tau^+\tau^-\tau^+\tau^-$ for a very light $A_1$ ($m_{A_1}< 2m_b$). This mode received recent attention as a potential probe for light pseudoscalar Higgs bosons in the NMSSM \cite{Dermisek:2010mg}, which spurred a reanalysis of old data and the publication of first experimental results in the interesting mass range by ALEPH \cite{Schael:2010aw}. These new results are included in our analysis. 

In addition to the neutral Higgs searches, the LEP experiments also published negative searches for charged Higgs bosons in the pair production channel $e^+e^-\to H^+H^-$, followed by the decays $H^\pm\to\tau^+\nu_\tau$, $H^\pm\to cs$, or $H^\pm\to W^\pm A$ in all the different combinations \cite{Abdallah:2003wd}. The latter mode is not relevant in the MSSM, but could become important in the NMSSM (or a general two-Higgs-doublet model) with a sufficiently light $A_1$. The resulting direct limits on $m_{H^\pm}$ depend slightly on the precise branching ratios, but in general the combined lower bound is around $m_{H^\pm}>80$--$90 \GeV$. We use $m_{H^\pm}>m_W$ as the lower limit for our NMSSM scans in this study. The precise charged Higgs boson limits are taken into account using \HB.

\subsubsection{The Tevatron}
The two Tevatron experiments CDF and D{\O} are actively searching for Higgs bosons in $p\overline{p}$ collisions at $1.96\TeV$, and their combined results currently exclude a SM-like Higgs at $2\,\sigma$ in the mass interval $158\GeV < m_h < 175\GeV$ \cite{TevHiggs:2010ar}. This range is already beyond the theoretically accessible region in the models we consider, but for supersymmetric Higgs bosons with enhanced couplings to fermions, new modes that are currently not excluding for the SM Higgs become interesting. In particular the search for $A$ production, followed by the decay $A\to \tau^+\tau^-$, provides strong exclusion for high $\tan\beta$ due to the enhancement of the $A$ coupling to down-type fermions. 

In addition to the neutral Higgs results, mass limits exist for a charged Higgs boson decaying in the $H^\pm\to \tau^\pm\nu_\tau$ and $H^\pm\to cs$ modes from CDF \cite{Aaltonen:2009ke} and D{\O} \cite{Abazov:2009zh}. In the case of a light charged Higgs boson ($m_{H^\pm}<m_t$) the limits are reported on $\mathrm{BR}(t\to bH^\pm)$, assuming either of the two modes saturates the $H^\pm$
width. In the NMSSM with a light $A_1$---where $H^\pm\to W^\pm A_1$ can be relevant---the experimental results can be conservatively interpreted as upper limits on the rescaled quantities $\xi_{H^\pm cs}\equiv
\mathrm{BR}(t\to bH^\pm)\times \mathrm{BR}(H^\pm\to c\overline{s})$ and
$\xi_{H^\pm\tau\nu}\equiv \mathrm{BR}(t\to bH^\pm)\times \mathrm{BR}(H^\pm\to
\tau^\pm\nu_\tau)$. An additional constraint can be derived from the search for $H^\pm\to W^\pm A_1$, $A_1\to \tau^+\tau^-$. This channel was recently analysed for the first time \cite{CDFWA:2010}.

Results from both neutral and charged Higgs bosons searches in the main channels at the Tevatron are incorporated in the \HB\ code. We use this to evaluate constraints both for the MSSM and the NMSSM. The charged Higgs constraints are cross-checked with an independent implementation where we also include the $H^\pm\to W^\pm A_1$ mode.

\subsubsection{Combined results}

\begin{figure}[h!]
\centering
\includegraphics[width=0.4\columnwidth]{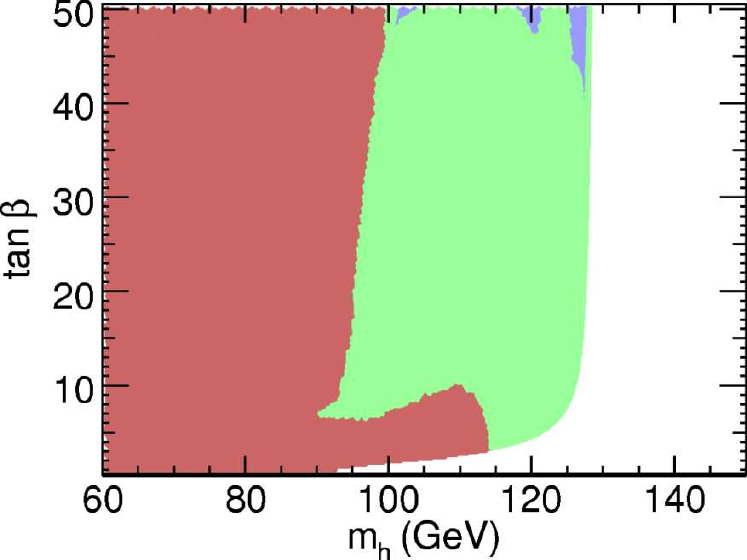}\;\;\;
\includegraphics[width=0.4\columnwidth]{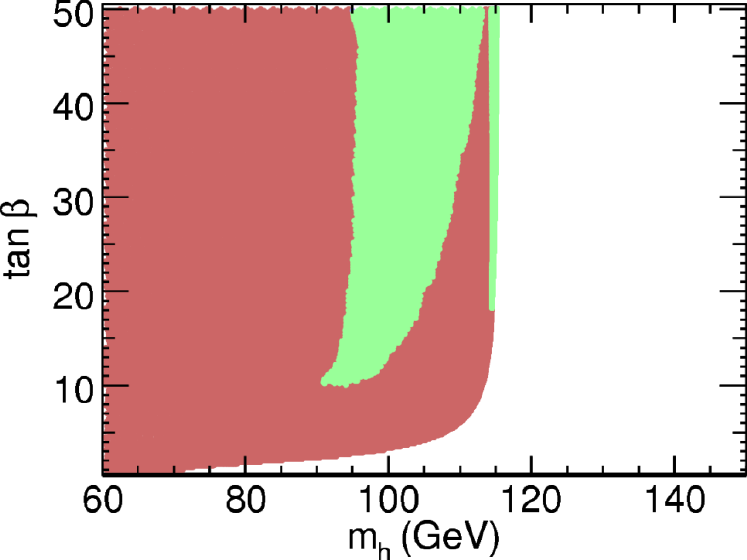}
\includegraphics[width=0.4\columnwidth]{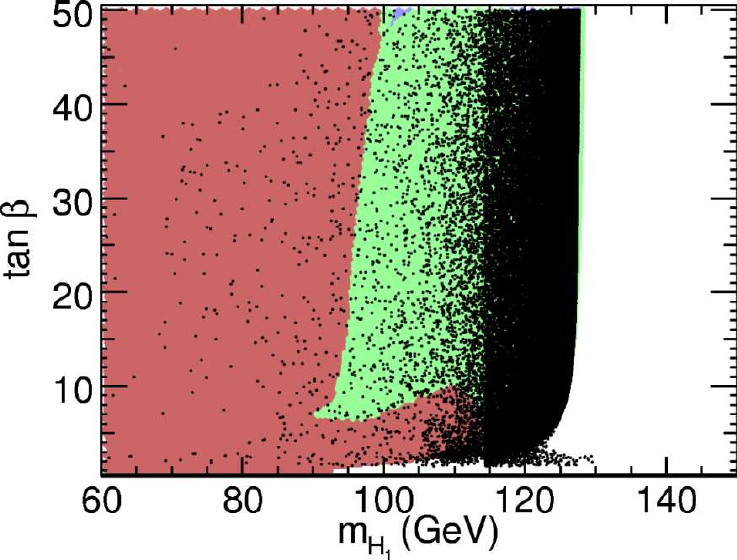}\;\;\;
\includegraphics[width=0.4\columnwidth]{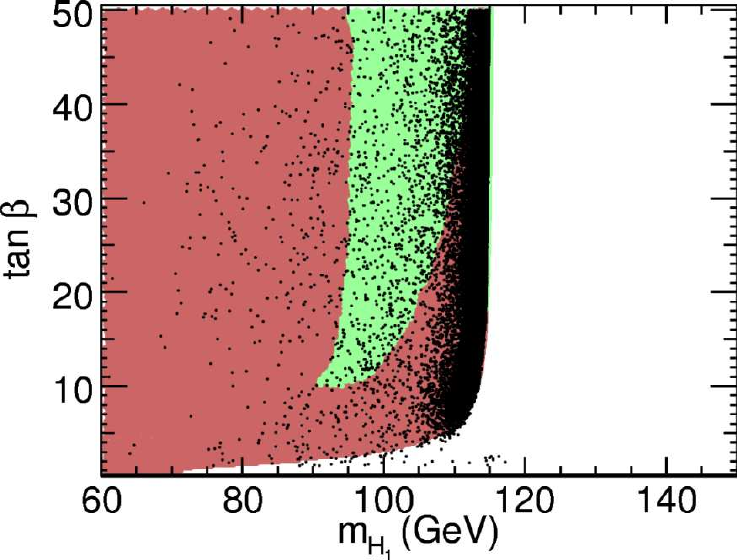}
\caption{Combined collider constraints in the $(m_{H_1},\tan\beta)$ plane for scenario (A) (left), and scenario (B) (right). The coloured regions represent exclusion in the MSSM at $95\%$ CL from LEP (red), the Tevatron (blue), and allowed parameter space (green). The black points represent NMSSM models which are compatible with all the collider constraints at $95\%$ CL.}
\label{fig:tbmh}
\end{figure}

We combine MSSM and NMSSM constraints from LEP and the Tevatron in each figure. This is demonstrated in Figure~\ref{fig:tbmh}, which shows the lightest CP-even Higgs mass versus $\tan\beta$. The solid (coloured) regions are results obtained for the underlying MSSM scenario: the red (blue) areas are excluded by LEP (Tevatron) at $95\%$ CL, while the green regions are allowed. White areas are theoretically inaccessible in the MSSM scenario. The scattered points (in black) are superimposed on the MSSM regions. They are the result of the NMSSM scans, with each point representing a choice of parameters compatible with all the collider constraints. 

Considering first scenario (A) (left two plots), a number of interesting features can be seen. There is very good agreement for a wide range of $\tan\beta$ between the upper limits on $m_{H_1}$ in the MSSM (shown in the top row) and the NMSSM (lower). This demonstrates that the dominant radiative corrections are taken into account consistently, and that they indeed are similar. The upper limit on $m_{H_1}$ is higher in the NMSSM in a narrow band at low $\tan\beta\sim 2.5$, where the tree-level $m_{H_1}$ is maximized by additional NMSSM contributions \cite{Drees:1998pw}. Focusing now on the excluded regions, it is clear from Figure~\ref{fig:tbmh} that the MSSM region of $\tan\beta\lesssim 10$, where essentially the SM limit $m_{h}>114\GeV$ applies, is reopened in the NMSSM. Additionally, the almost $\tan\beta$--independent limit of $m_h\gtrsim 90\GeV$ is not respected, but valid NMSSM points are observed down to $m_{H_1}\simeq 60\GeV$. That such light Higgs bosons avoid the collider constraints is due to a large singlet component of $H_1$, making its coupling to vector bosons smaller. In a situation where the $H_2$ has a reduced coupling with respect to the SM $C(H_2VV)^2\simeq 1$, while the lightest Higgs boson has $C(H_1VV)^2\simeq 0$, the lightest state $H_1$ is not produced through Higgs-strahlung and the LEP limits do not apply. (For a precise definition of the reduced couplings we refer to Table~\ref{tab:couplings} below.)

The situation in scenario (B) (right plot) appears somewhat different, mainly because of the modified MSSM conditions. Close to the maximum value $m_h\sim 115\GeV$, only a narrow strip where $\tan\beta>15$ is allowed in the MSSM. On the other hand, there is still an allowed region for somewhat lower $m_{H_1}$, ending at the decay mode independent LEP limit $m_h>90\GeV$, which remains similar to before. Even if the valid NMSSM points tend to cluster at large values for $m_{H_1}$, their existence in all the MSSM-excluded regions is evident. The region $110\GeV < m_{H_1} < 115\GeV$ is allowed because the decay $H_1\to A_1 A_1\to 4b$ is open, and the LEP constraints for this mode are weaker than for the $H_1\to b\bar{b}$ decay which is always dominant in the MSSM for this mass range. 
\begin{figure}[h!]
\centering
\includegraphics[width=0.4\columnwidth]{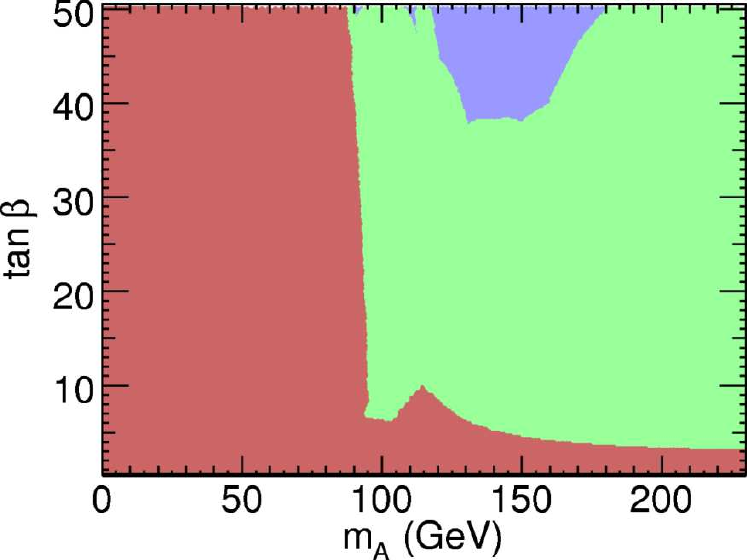}\;\;\;
\includegraphics[width=0.4\columnwidth]{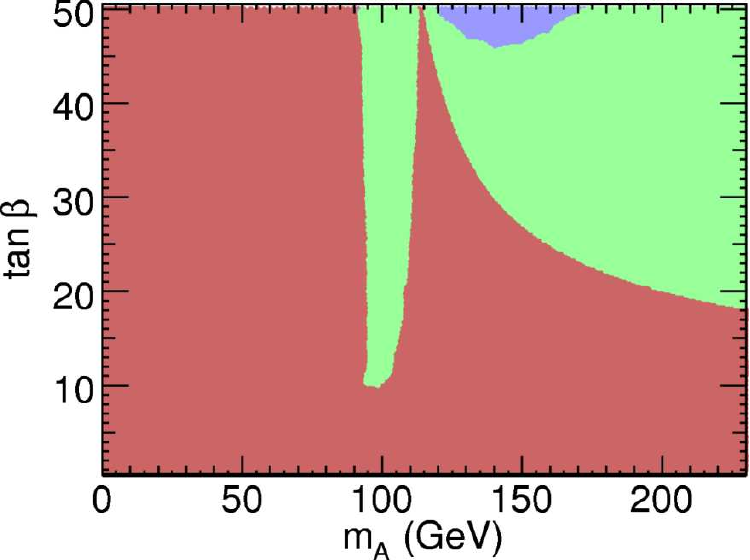}
\includegraphics[width=0.4\columnwidth]{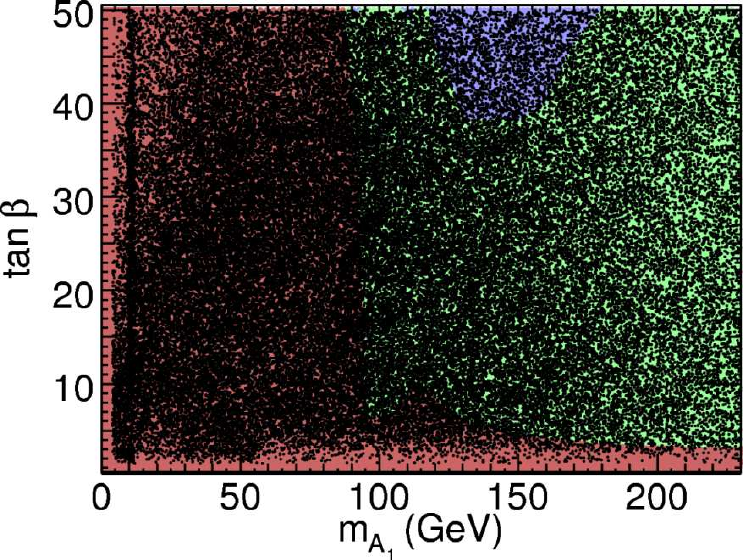}\;\;\;
\includegraphics[width=0.4\columnwidth]{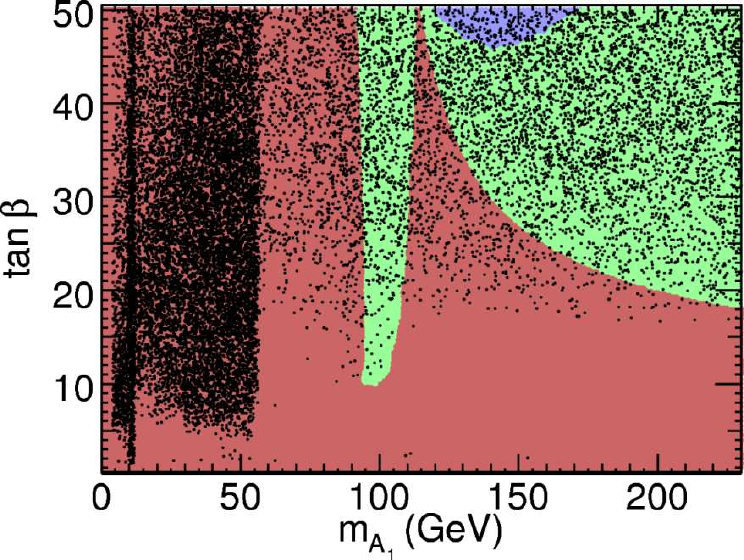}
\caption{Combined collider constraints in the $(m_{A_1},\tan\beta)$ plane for scenario (A) (left), and scenario (B) (right). The colour coding is the same as for Figure~\ref{fig:tbmh}.}
\label{fig:tbmA}
\end{figure}

For the lightest CP-odd Higgs boson $A_1$, exclusion plots of the same type as for $H_1$ is shown in Figure~\ref{fig:tbmA}. In scenario (A), a similar result as for $H_1$ with respect to the LEP constraints can be observed: the MSSM limit $m_A\gtrsim 90\GeV$ is not respected by $m_{A_1}$ in the NMSSM. The $90\GeV$ limit applies only to the effective doublet mass (and then only when $m_{A_1}>2 m_b$). We find points with a large doublet component also below the $b\bar{b}$ threshold. This supports the results of \cite{Dermisek:2008uu}, and shows that scenarios like this are viable with the updated collider constraints we have included. The most important constraint in this mass range is the new result from ALEPH on $H_1\to A_1A_1\to 4\tau$. It relates a light $A_1$ to specific requirements on the mass of $H_1$. Either $m_{H_1}>107\GeV$ and the constraint does not apply, or $H_2$ must be SM-like so that the coupling $C(H_1VV)$ of $H_1$ to vector bosons is suppressed.
In the intermediate range $2 m_b < m_{A_1} < 90\GeV$, the doublet mass $m_A$ has to be above $90\GeV$. These values for $m_{A_1}$ are allowed in the NMSSM since the reduced coupling $C(H_1A_1Z)$ can be lower than in the MSSM, which decreases the rate for associated $H_1A_1$ production. We find that $C(H_1A_1Z)^2\lesssim 0.5$ is sufficient to find allowed $m_{A_1}$ over the full range (see also Figure~\ref{fig:phiVV} below).
Having $m_{A_1}\ll 90\GeV$ of course opens interesting phenomenological possibilities, as will be discussed further in Section~\ref{sect:LHC}. Another feature which is clearly visible in Figure~\ref{fig:tbmA} is the Tevatron MSSM exclusion at high $\tan\beta$ and intermediate $m_A$. This exclusion does not remain valid in the NMSSM when the relevant couplings $C(A_1b\bar{b})$, $C(A_1\tau^+\tau^-)\propto \tan\beta \cos\theta_A$ are suppressed by singlet mixing ($|\cos\theta_A|<1$).

In scenario (B), the allowed NMSSM points divide into distinct classes. For $m_{A_1}<2m_b$, the same possibility as discussed for scenario (A) exists to have an $A_1$ with a large doublet component. The same conditions on the mass of $H_1$ as discussed for scenario (A) apply also in this case. In the region $2m_b<m_{A_1}<57\GeV$ we find points which are associated with a dominant decay $H_1\to A_1A_1$. As discussed above, a substantial branching fraction for this mode circumvents limits on $m_{H_1}$, which leads to the large number of allowed points in this mass range. The upper limit is set by the kinematics from the upper limit on $m_{H_1}$ in scenario (B). In addition to these two classes, there is a sparse population of scattered points over most of the plane, preferentially in the MSSM-allowed areas. Where necessary, these avoid the collider limits by a small coupling $C(H_1A_1Z)$, as also discussed for scenario (A). We note that $\tan\beta<10$ is disfavoured for $m_{A_1}>57\GeV$ also in the NMSSM, since in this region it is difficult to find allowed values for $m_{H_1}$. The only exception is for $\tan\beta\simeq 2.5$ where the tree-level value for $m_{H_1}$ is maximal.

\begin{figure}
\centering
\includegraphics[width=0.48\columnwidth]{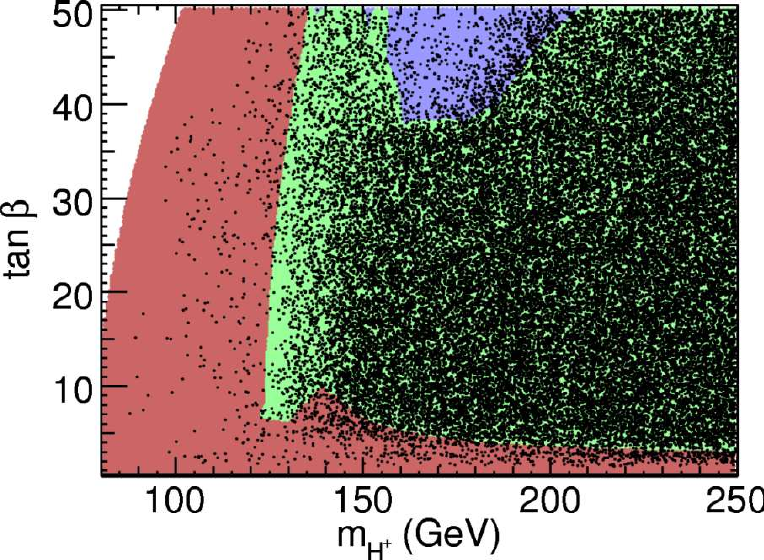}\;\;\;
\includegraphics[width=0.48\columnwidth]{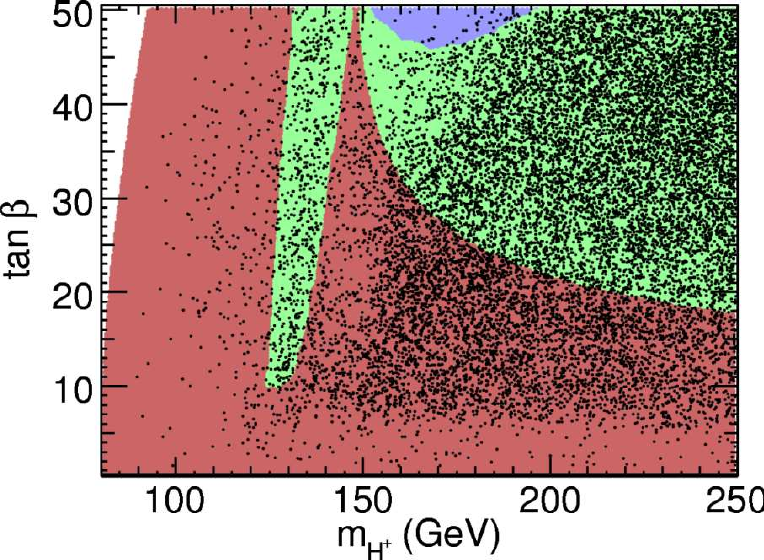}
\caption{Combined collider constraints in the $(m_{H^\pm},\tan\beta)$ plane for scenario (A) (left), and scenario (B) (right). The colour coding is the same as for Figure~\ref{fig:tbmh}. Constraints from direct searches for $H^\pm$, as well as indirect constraints from neutral Higgs searches, are included.}
\label{fig:tbmhc}
\end{figure}

Figure~\ref{fig:tbmhc} shows the combined constraints in the plane of the charged Higgs mass and $\tan\beta$. In the MSSM, these are quite similar to the results for the CP-odd scalar. This is a consequence of the tree-level mass relation $m_{H^\pm}^2=m_A^2+m_W^2$, which typically receives only small corrections at higher orders. As a result, an indirect limit $m_{H^\pm}>120\GeV$ can be derived from the previous result $m_A>90\GeV$. Turning now to the NMSSM results, it can be seen that this limit does not apply. There are two reasons why it is not effective: first---as we have already seen---even a doublet-like $A_1$ can be very light ($m_{A_1}<2 m_b$, which can correspond to an effective doublet mass $m_A< 90\GeV$). Second, and equally important, the tree-level mass relation is modified, cf.~Equation~\eqref{eq:mHc}. For this reason, we find the largest number of points with light $H^\pm$ for large values of the NMSSM coupling $\lambda$. Paradoxically, the MSSM limit ($\lambda\to 0$) is thus most clearly visible in the charged Higgs constraints---even though no new charged degrees of freedom have been introduced. In addition to the indirect constraints from the neutral Higgs searches, which are most excluding, $m_{H^\pm}$ is limited by the direct search results. Here the decay $H^\pm\to W^\pm A_1$ can degrade the standard search channels ($H^\pm\to \tau^\pm\nu_\tau$, $H^\pm\to cs$), which would also make it easier to accommodate a light charged Higgs. However, we find that with the present (limited) reach of the Tevatron, it is not necessary to invoke this mode to avoid exclusion, save for very high (or very low) $\tan\beta$. We expect this to change quite dramatically when results from the LHC become available. As a final remark on Figure~\ref{fig:tbmhc}, we would like to point out that the low $\tan\beta$ region is not excluded in the NMSSM. It could therefore be potentially interesting to search for two jet decays of $H^\pm$. The situation for the charged Higgs boson in scenario (B) is very similar to scenario (A). Also in this case there exist points with $m_{H^\pm}\sim 90\GeV$ which are still allowed. We note that a low $\tan\beta$ is only possible when the $A_1$ is comparably light, cf.~Figure~\ref{fig:tbmA}, which means the decay $H^\pm\to W^\pm A_1$ is likely to be dominant for $\tan\beta<10$ in this scenario.

\subsection{Flavour physics}
In addition to the collider constraints, we consider limits resulting from several flavour physics processes on the two benchmark scenarios (A) and (B). The same observables have been applied previously to constrain the Higgs sector in the MSSM in the context of the CMSSM and NUHM frameworks \cite{Barenboim:2007sk,Eriksson:2008cx,Akeroyd:2009tn}, and a subset of them also in the NMSSM \cite{Domingo:2007dx}. For the theoretical evaluation of the model predictions, we use the code \SI\ {\tt v3.0} \cite{Mahmoudi:2007vz,*Mahmoudi:2008tp,Mahmoudi:2009zz} 
for the calculations both in the MSSM and the NMSSM. \SI\ provides automatic interfaces with several MSSM and NMSSM spectrum generators. Here we used {\tt SOFTSUSY 3.1.6} \cite{Allanach:2001kg} and {\tt SuSpect 2.41} \cite{Djouadi:2002ze} for the MSSM and {\tt NMSSMTools 2.2.0} \cite{Ellwanger:2005dv,*Ellwanger:2004xm} for the NMSSM.
\begin{table*}
\centering
\begin{tabular}{lrcr}
\hline
Observable & Experimental value & & $95\%$ CL range\\
\hline
$\displaystyle \mathrm{BR}(B\to X_s\gamma)$ & $(3.55\pm 0.24\pm 0.09)\times 10^{-4}$ & \cite{Barberio:2008fa} & $ \left[2.16,4.93\right]\times 10^{-4}$ \\
$\displaystyle \mathrm{BR}(B_s\to \mu^+\mu^-)$ & $< 4.3\times 10^{-8}$ &
\cite{CDFBsmumu:2009} & $ < 4.7\times 10^{-8} $\\
$\mathrm{BR}(B\to \tau\nu_\tau)/\mathrm{BR}(B\to \tau\nu_\tau)^{\mathrm{SM}}$ & $1.63\pm 0.54$ & \cite{Barberio:2008fa} & $ \left[0.56,2.70\right]$  \\
$\mathrm{BR}(B\to D\tau\nu_\tau)/\mathrm{BR}(B\to De\nu_e)$ & $0.416\pm 0.117\pm 0.052$ & \cite{Aubert:2007dsa} & $ \left[0.151, 0.681\right]$ \\
$R_{\ell 23}(K\to\mu\nu)$ & $1.004\pm 0.007$ & \cite{Antonelli:2008jg} & $ \left[0.990,1.018\right]$\\
$\mathrm{BR}(D_s\to\tau\nu_\tau)$ & $(5.38\pm 0.32)\times 10^{-2}$ & \cite{Barberio:2008fa} & $ \left[4.7, 6.1\right]\times 10^{-2}$ \\
\hline
\end{tabular}
\caption{Experimental values for flavour physics observables and the corresponding allowed ranges at $95\%$ confidence level.}
\label{tab:flav}
\end{table*}

To determine the allowed intervals for the model to be compatible with the data at $95\%$ CL, in the presence of both experimental and theoretical uncertainties, a Monte Carlo approach is used \cite{Statistics}. The experimental values used as input for our analysis and the resulting allowed ranges are given in Table~\ref{tab:flav}. Since the main objective is to compare the two models, no attempt is made at a sophisticated statistical combination of the different constraints or a global fit. Instead we simply give the exclusion at $95\%$ CL by each observable separately, meaning that any particular point may be excluded by more than one measurement. This approach should be sufficient to give a general indication of what the viable regions for the Higgs masses in the NMSSM are.

The NMSSM specific additions to $\rm{BR}(B \to X_s \gamma)$ required to take into account the additional contributions from the Higgs bosons and the extra neutralino are obtained according to a generalisation of \cite{Buras:2002vd,*Hofer:2009xb,Heng:2008rc}. The NMSSM effects in this decay appear only at the two-loop level and therefore do not lead to substantial departure from the MSSM results.

The implementation of BR($B_s \to \mu^+ \mu^-$) is based on \cite{Heng:2008rc,Bobeth:1999mk,*Bobeth:2001jm,*Bobeth:2004jz,Hiller:2004ii} and is in the approximation of large $\tan\beta$. Contrary to BR($B \to X_s \gamma$), here one can expect some differences with respect to the MSSM. Indeed, when $A_1$ is light (which can be possible for any $m_{H^\pm}$), it generates sizeable contributions to the branching ratio through penguin diagrams, especially at large $\tan\beta$. On the other hand, even for large $m_{A_1}$ the coupling of $A_1$ with down type quarks can take both small and large values \cite{Hiller:2004ii}, and if this coupling is large the $A_1$ contribution can become competitive or even dominant. For the numerical evaluations we use a value for the $B_s$ decay constant of $f_{B_s}=238.8\pm 9.5$ MeV \cite{Laiho:2009eu}.

In the leptonic (and semi-leptonic) decays of pseudoscalar mesons, the dominating contribution beyond the SM is through charged Higgs exchange diagrams. The calculation of these observables have been simply extended from the usual MSSM results to the NMSSM. 
Since this sector is not directly modified in the NMSSM, the impact on flavour physics is only indirect. Small differences can arise due to the fact that the charged Higgs mass in the NMSSM can be lower than in the MSSM for a similar neutral Higgs spectrum. Here we consider the branching ratios of $B\to \tau\nu_\tau$, $B\to D\tau\nu_\tau$, $D_s \to \tau\nu_\tau$ and $K\to \mu \nu$, and we use $|V_{ub}|=(3.92\pm0.09\pm0.45)\times10^{-3}$ \cite{Charles:2004jd}, 
$f_B=192.8\pm9.9$ MeV \cite{Laiho:2009eu}, $f_{D_s}= 248 \pm 2.5$ MeV \cite{Davies:2010ip}, and $f_K/f_\pi = 1.189 \pm 0.007$ \cite{Follana:2007uv} for the numerical evaluations. 

All the formulae and the detail of the calculations of the above branching ratios can be found in \cite{Mahmoudi:2007vz,*Mahmoudi:2008tp}. 

In addition to the aforementioned observables, we apply also constraints from $\Upsilon(1S) \to A_1 \gamma$ \cite{Domingo:2008rr} as implemented in {\tt NMSSMTools 2.2.0}. This excludes many points with a light $A_1$ in the range $4\GeV < m_{A_1}< 9\GeV$. However, since this observable does not exclude a distinct region in ($m_{H^\pm},\tan\beta$), the constraints are not shown with a separate colour in the figures (though points termed `allowed' always fulfil this constraint). We refrain from applying the new constraints from $\Upsilon(3S) \to A_1 \gamma$ decay \cite{Aubert:2009cka,Aubert:2009cp,Dermisek:2010mg} since theoretical uncertainties in the calculation of the exclusive Upsilon decays are not well under control \cite{SanchisLozano:2010ij}. 
\begin{figure}
\centering
\includegraphics[width=0.45\columnwidth]{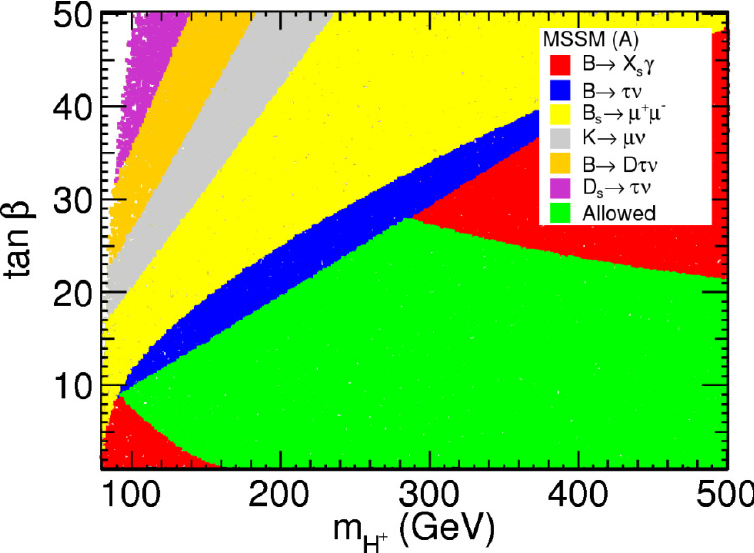}\;\;\;
\includegraphics[width=0.45\columnwidth]{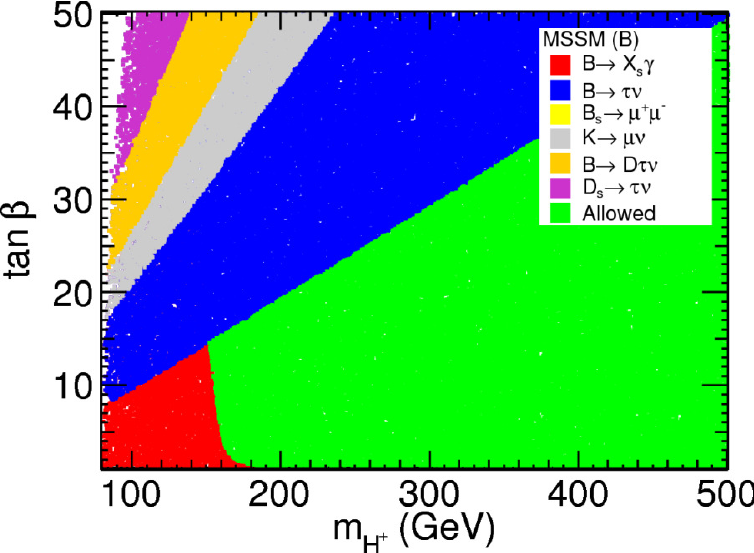}
\includegraphics[width=0.45\columnwidth]{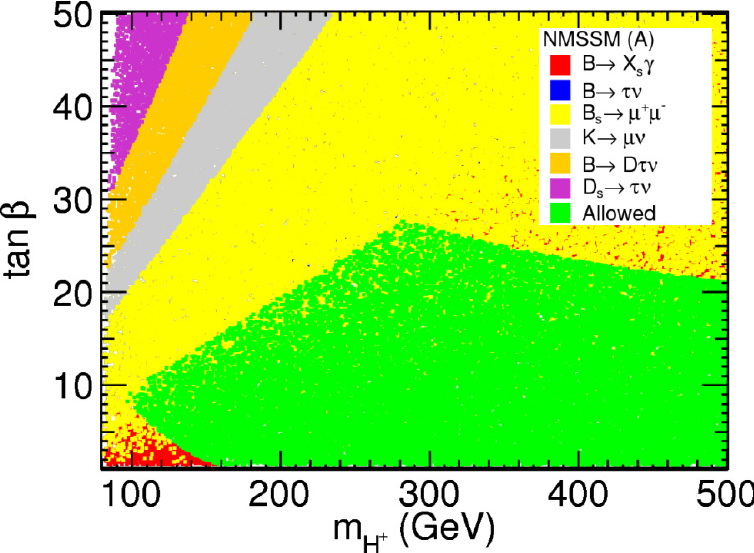}\;\;\;
\includegraphics[width=0.45\columnwidth]{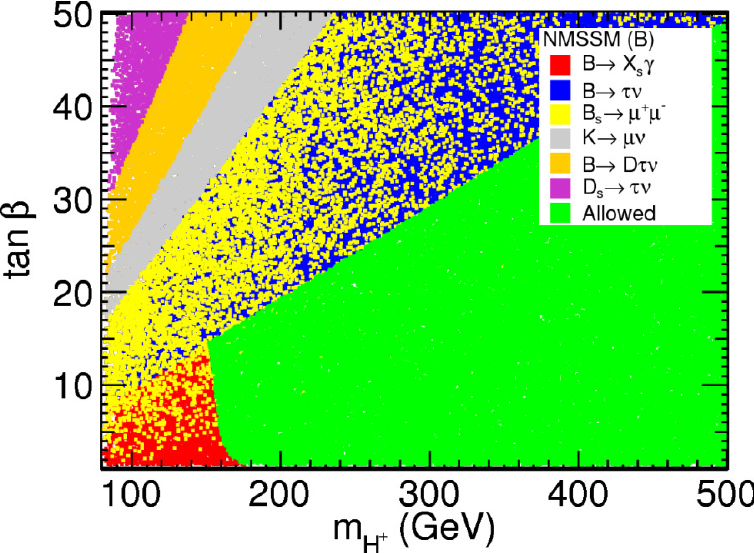}
\caption{Flavour constraints on the $(m_{H^\pm},\tan\beta)$ plane in the MSSM (upper) and the NMSSM (lower) for scenario (A) (left) and scenario (B) (right). The constraints are applied in the order indicated by the legend with the points satisfying all constraints in green in the foreground.}
\label{fig:flav_cmhmax}
\end{figure}

In Figure~\ref{fig:flav_cmhmax} we show a comparison between the flavour constraints in scenario (A) (to the left) for the MSSM (upper row) and the NMSSM (lower row). We find it most useful to display the exclusion in the plane $(m_{H^\pm}$, $\tan\beta)$. We notice the large similarities between the MSSM and the NMSSM allowed regions. Indeed, even if the $B_s \to \mu^+ \mu^-$ branching ratio can be different in the NMSSM, the allowed region keeps a similar shape, since the extra parameters of the NMSSM enable many of the points to evade the $B_s \to \mu^+ \mu^-$ constraint. Allowed points are found down to $m_{H^\pm}\sim 95\GeV$ ($100\GeV$) for the MSSM (NMSSM).
The corresponding results for scenario (B) are given in the right column of Figure~\ref{fig:flav_cmhmax}. The allowed region is again similar in both the MSSM and the NMSSM, and is larger than in scenario (A) since the lower limit on $B\to X_s\gamma$ does not come into play. On the other hand we note that the limit on $m_{H^\pm}$ is stronger, $m_{H^\pm}\gtrsim 145\GeV$ in both cases. We also note that the $B_s \to \mu^+ \mu^-$ branching ratio is less constraining in scenario (B).

\begin{figure}
\centering
\includegraphics[width=0.45\columnwidth]{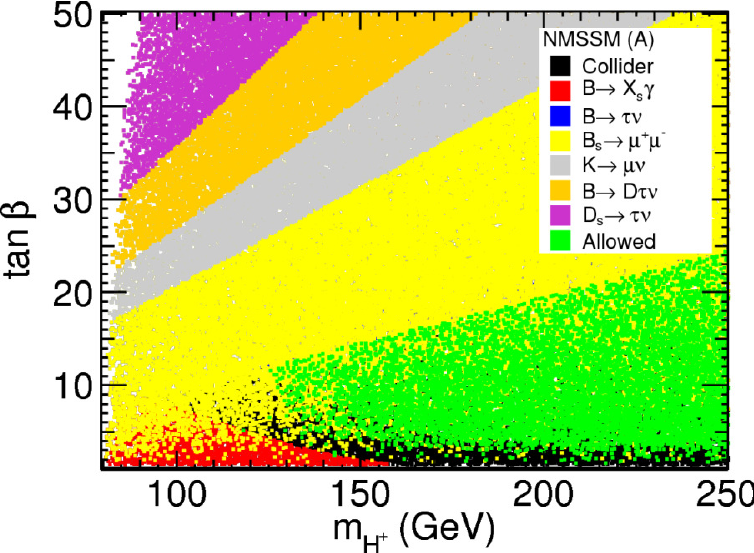}\;\;\;
\includegraphics[width=0.45\columnwidth]{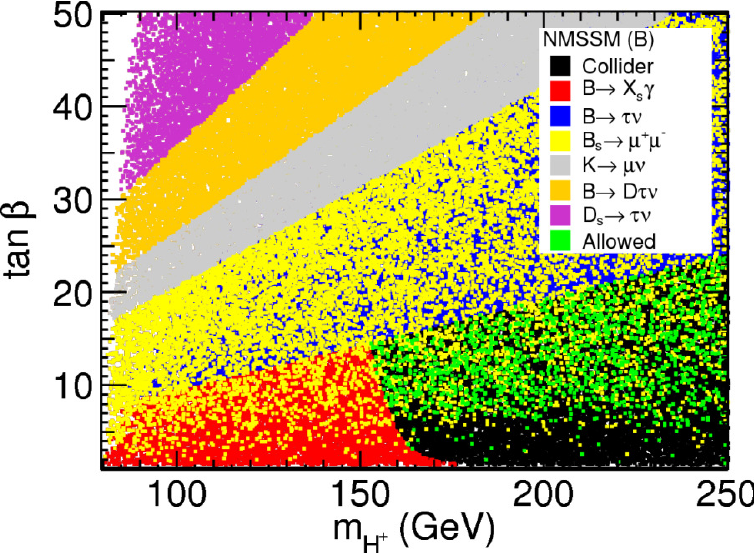}
\caption{Combined constraints from collider and flavour physics in the NMSSM for scenario (A) (left) and scenario (B) (right). The constraints are applied in the order indicated by the legend, with the allowed points on top. }
\label{fig:flav_all}
\end{figure}

In Figure~\ref{fig:flav_all}, we combine the flavour constraints with the constraints from colliders presented in the previous section. When all constraints are applied, we do no longer observe any accepted points with $m_{H^\pm}<120\GeV$.

\begin{figure}
\centering
\includegraphics[width=0.45\columnwidth]{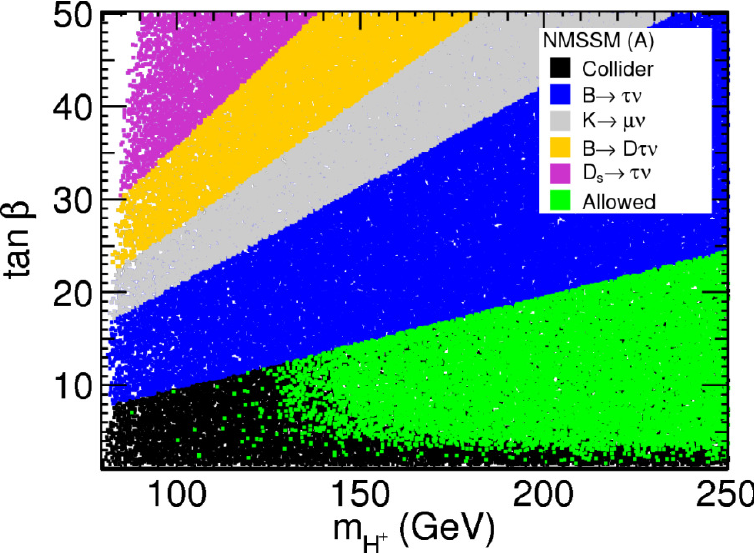}\;\;\;
\includegraphics[width=0.45\columnwidth]{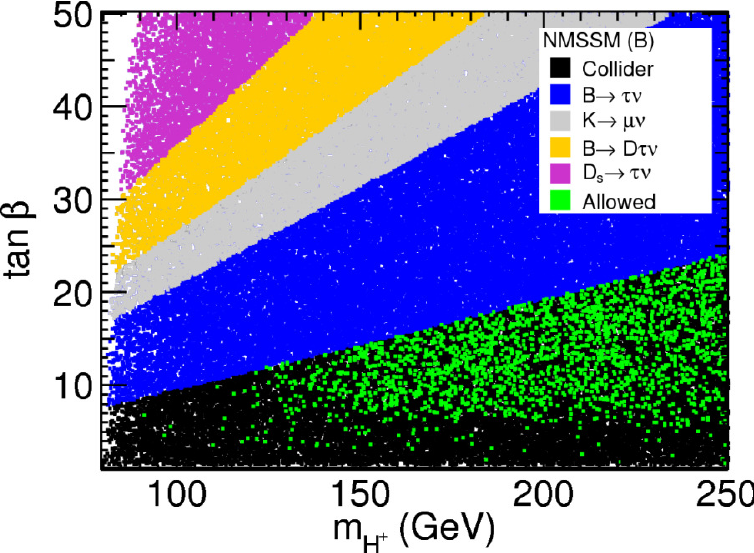}
\caption{Combined constraints from collider and flavour physics in the NMSSM for scenario (A) (left) and scenario (B) (right). Only flavour observables with tree-level $H^\pm$ exchange are included. The constraints are applied in the order indicated by the legend, with the allowed points on top.}
\label{fig:flav_all_tree}
\end{figure}

Knowing that the constraints from $B\to X_s\gamma$ and $B_s\to\mu^+\mu^-$ depend on the choice of SUSY scenario (as can clearly be seen by comparing the results for our two scenarios), it is also motivated to look at the situation when these constraints are removed. This is presented in Figure~\ref{fig:flav_all_tree}, which only includes the direct constraints from colliders and those flavour physics constraints which are mediated by $H^\pm$ at tree-level. As already discussed, these are robust with respect to a change of the SUSY scenario. When only these `unavoidable' constraints are applied, the situation at low $\tan\beta\lesssim 10$ is once again determined by the collider constraints. Hence the results from the previous section applies in this region, and we find points with $m_{H^\pm}\lesssim 100\GeV$ around $\tan\beta=5$. The high $\tan\beta$, low $m_{H^\pm}$ region on the other hand is strongly disfavoured by flavour constraints (both in the MSSM and the NMSSM).

One constraint which was not discussed so far is that from the anomalous magnetic moment of the muon, $a_\mu=(g_\mu-2)/2$. As is well-known, the SM calculation of this
quantity disagrees with the measurement of the E821 experiment. The discrepancy currently amounts to a $3.6\,\sigma$ effect \cite{Davier:2010nc} when hadronic contributions are evaluated with
data from $e^+e^-$ collisions, corresponding to $a_\mu^{\mathrm{exp}}-a_\mu^{\mathrm{SM}}=(28.7\pm 8.0)\times 10^{-10}$. A slightly smaller deviation is observed when $\tau$ decay data is used instead. A positive contribution from supersymmetry (most importantly from chargino/sneutrino or neutralino/smuon loops) can explain this deviation. However, this requires that the sign of the $\mu$ parameter is positive, which in the NMSSM translates into the requirement $\mu_{\mathrm{eff}}>0$. Even with this choice, the SUSY mass scale in the benchmark scenarios employed here is too high to explain the full deviation, but the difference between data and theory is reduced. It would be possible to explain the full difference by departing from scalar mass universality and allow for a lower smuon mass. This would not have an important impact on any of the other processes we consider. The numerical differences we obtain in $a_\mu$ between the MSSM and the NMSSM is typically below $1\times 10^{-10}$, hence much smaller than the theoretical error and therefore unlikely to be interesting.

\section{LHC prospects}
\label{sect:LHC}

\begin{table}[h!]
\centering
\begin{tabular}{cccc}
\hline
\hspace{10pt}Vertex\hspace{10pt} & \hspace{10pt}NMSSM\hspace{10pt} & \hspace{10pt}MSSM\hspace{10pt} &
\hspace{10pt}SM\hspace{10pt}   \\
\hline
$H_1tt$ & $\dfrac{S_{11}}{\sin\beta}$ & $\dfrac{\cos\alpha}{\sin\beta}$ & 1 \\
$H_1bb$ & $\dfrac{S_{12}}{\cos\beta}$ & $\dfrac{\sin\alpha}{\cos\beta}$ & 1 \\
$H_1VV$ & $ \sin\beta S_{11} + \cos\beta S_{12}$ & $\sin(\beta-\alpha)$ & 1 \\
$H_2tt$ & $\dfrac{S_{21}}{\sin\beta}$ & $\dfrac{\sin\alpha}{\sin\beta}$ & {\rm n.a.} \\
$H_2bb$ & $\dfrac{S_{22}}{\cos\beta}$ & $\dfrac{\cos\alpha}{\cos\beta}$ & {\rm n.a.} \\
$H_2VV$ & $ \sin\beta S_{21} + \cos\beta S_{22}$ & $\cos(\beta-\alpha)$ & {\rm n.a.} \\
$A_1tt$ & $\cot\beta\cos\theta_A$ & $\cot\beta$ & {\rm n.a.} \\
$A_1bb$ & $\tan\beta\cos\theta_A$ & $\tan\beta$ & {\rm n.a.} \\
$A_2tt$ & $\cot\beta\sin\theta_A$ & ${\rm n.a.}$ & {\rm n.a.} \\
$A_2bb$ & $\tan\beta\sin\theta_A$ & ${\rm n.a.}$ &{\rm n.a.} \\ 
$A_1H_1Z$ & $ (\cos\beta S_{11} - \sin\beta S_{12})\cos\theta_A$   & $\cos(\beta-\alpha)$ &  {\rm n.a.} \\
$A_1H_2Z$ & $ (\cos\beta S_{21} - \sin\beta S_{22})\cos\theta_A$   & $\sin(\beta-\alpha)$ &  {\rm n.a.} \\
\hline
$H_1H^+W^-$ & $ \cos\beta S_{11} - \sin\beta S_{12}$ & $\cos(\beta-\alpha)$ & {\rm n.a.} \\
$A_1H^+W^-$ & $ \cos\theta_A$   & $1$ &  {\rm n.a.} \\
\hline
\end{tabular}
\caption{Reduced Higgs couplings in the NMSSM compared to the MSSM and the SM (when applicable). The couplings are defined such that the cross section for single Higgs and associated $A_i H_j$ production can be obtained by multiplying the relevant cross section in Figure~\ref{fig:sigma_h_SM} with the reduced coupling squared. It should also be noted that the reduced couplings to fermions are identical for all three generations, even if only the third generation is displayed here.}
\label{tab:couplings}
\end{table}

Given the higher centre of mass energy of the LHC compared to the Tevatron, the prospects for discovering NMSSM Higgs bosons is certainly much brighter at the LHC, but still remains a difficult task. In the following we will highlight some special features of the scenarios considered here that are relevant for the searches at the LHC. We discuss how they differ from the corresponding MSSM scenarios, as well as the impact of collider and flavour constraints. We will concentrate on the design energy for the LHC of 14 TeV, although many of the results we present are equally valid for lower centre of mass energies such as 7 or 8 TeV.

We start by considering the mass spectra of the two scenarios. As already alluded to the Higgs masses can differ substantially from the corresponding MSSM scenarios due to mixing with the additional singlet, as well as loosening of the collider and flavour constraints as discussed in the previous section. As a main result the lightest CP-even (and CP-odd) Higgs, as well as the charged Higgs, can be  lighter than in the corresponding MSSM scenarios. The relation of their masses to those of the remaining states $A_2$, $H_2$, and $H_3$ in general depends on which states originate from the first doublet, the second doublet, and from the singlet. The CP-odd Higgses can be classified as either doublet-like or singlet-like, while the CP-even doublet states can be further separated into SM-like or MSSM-like.

The CP-odd Higgs masses  are related to the doublet mass $m_A^2$ as $m^2_{A_1}\lesssim m^2_A$ and $m^2_{A_2}\gtrsim m_A^2$, where the near equality holds for a purely doublet state. Through Equation~\eqref{eq:mHc} they are similarly related to $m^2_{H^\pm}$. The mass of the singlet-like component is limited by $v_s$  which varies inversely with $\lambda$.
The CP-even Higgs spectrum is slightly more complicated. Recall that in the MSSM there is one SM-like state with tree-level mass $m_h \lesssim m_Z$ and one MSSM-like with mass  $m_H^2 \approx m_A^2=m_{H^\pm}^2-m_W^2$. In the NMSSM, the mixing with the singlet state (with mass $m_S \lesssim v_s$) will then give three states with masses $m_{H_3}^2 \gtrsim \max{(m_h,m_H)}$, 
 $\min{(m_h,m_H)} \lesssim m_{H_2}^2 \lesssim \max{(m_h,m_H)}$, and
 $m_{H_1}^2 \lesssim \min{(m_h,m_H)}$. The (upper) equalities again holds for the pure SM or MSSM-like states respectively, whereas the singlet-like state can have any mass.

In summary, the charged Higgs mass determines the overall features of the mass spectrum for the neutral Higgses through Equation~\eqref{eq:mHc}. For large $m_{H^\pm} \gtrsim v$, both
$A_1$ and $H_2$ will have masses that are smaller than or similar to the charged Higgs mass, whereas $A_2$ and $H_3$ will be heavier. Still, both $A_2$ and $H_3$ can have masses of phenomenological interest for the LHC. However, since $H_3$ does not add any essential new phenomena compared to $H_2$, we will not discuss it further in the following.

\begin{figure}
\centering
\includegraphics[width=0.48\columnwidth]{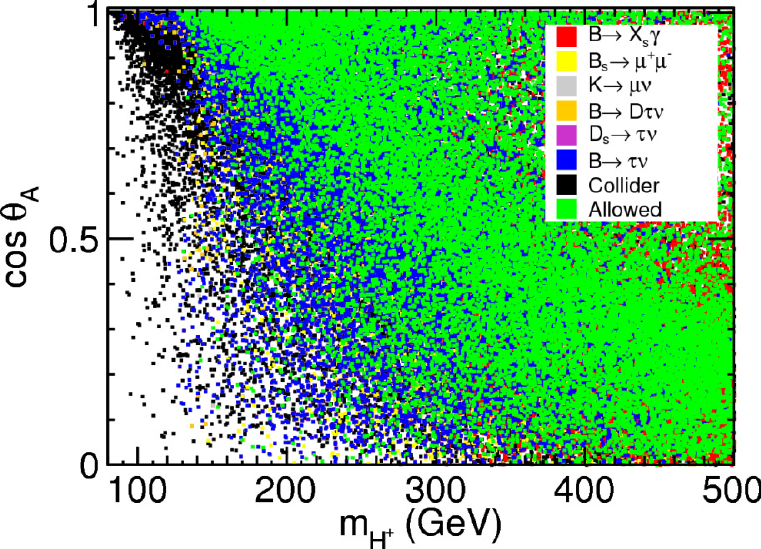}\;\;\;
\includegraphics[width=0.48\columnwidth]{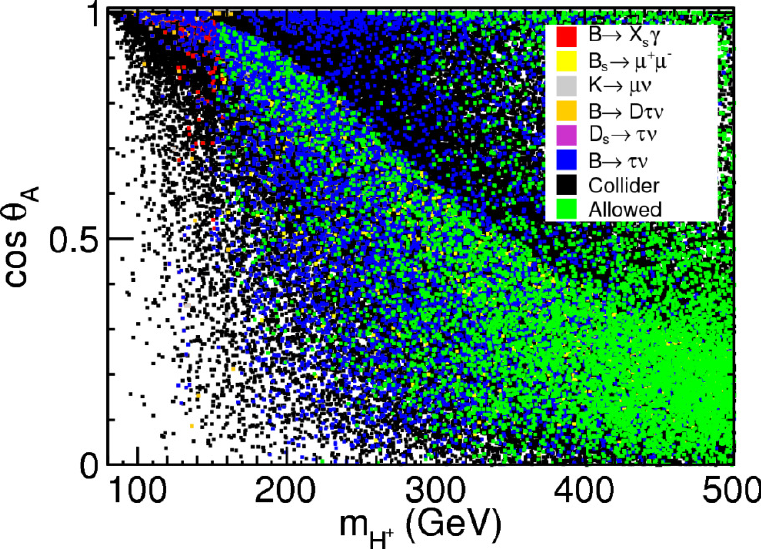}
\includegraphics[width=0.48\columnwidth]{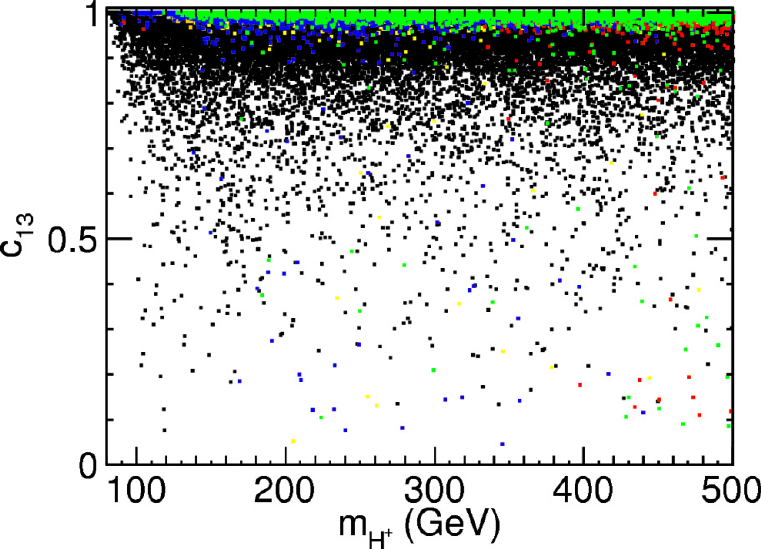}\;\;\; 
\includegraphics[width=0.48\columnwidth]{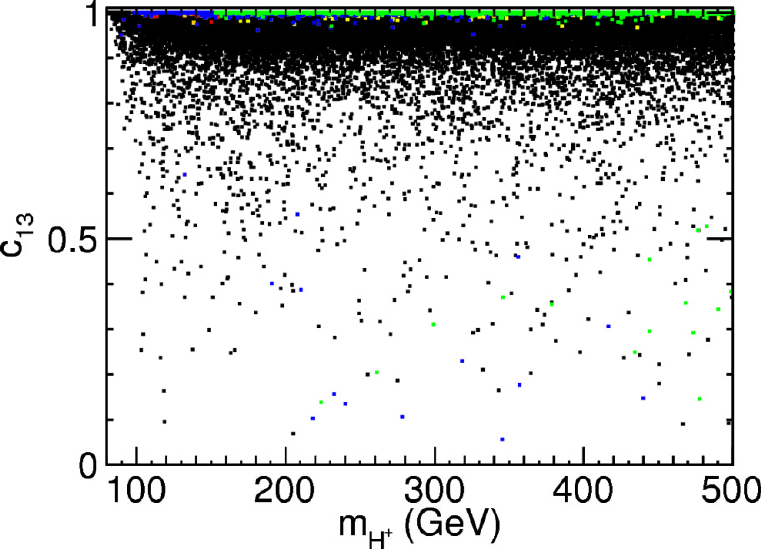}
\includegraphics[width=0.48\columnwidth]{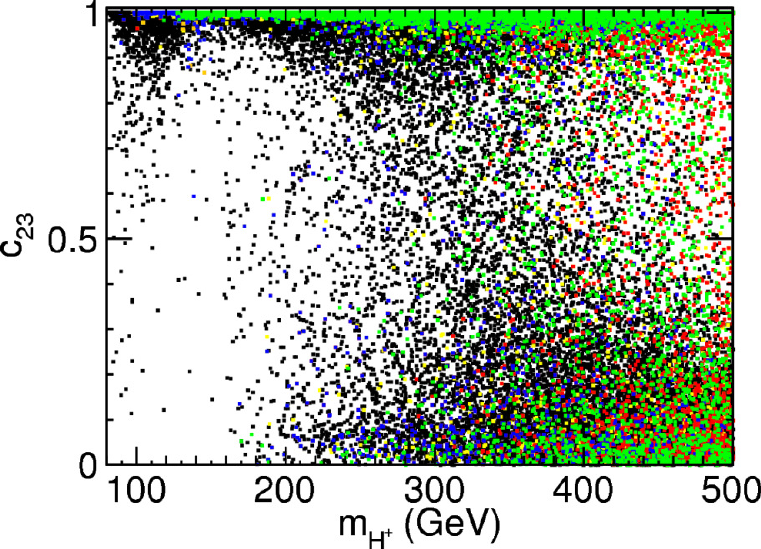}\;\;\; 
\includegraphics[width=0.48\columnwidth]{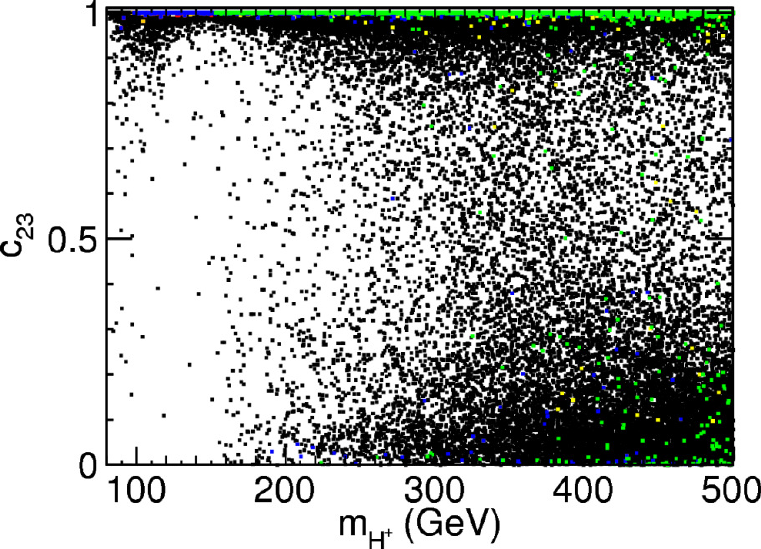}
\caption{The mixing angles $\cos\theta_A$, $\cos\theta_{13}$ and $\cos\theta_{23}$ shown as a function of the charged Higgs boson mass for scenario (A) (left) and scenario (B) (right). The constraints are applied in the order indicated by the legend.}
\label{fig:CosThetaA}
\end{figure}

Using the conventions introduced in Section~\ref{sect:Higgs} the couplings of the Higgs bosons to fermions and vector bosons, as well as the triple Higgs boson couplings, can be written down. In Table~\ref{tab:couplings} we give these couplings in their reduced form, meaning that common coupling constants, kinematic factors etc.~are left out and only the dependence on the mixing angles is kept. Thus the reduced couplings give a measure of the relative couplings of the NMSSM compared to the MSSM and the SM (where applicable). As is clear from Table~\ref{tab:couplings} the NMSSM couplings approach their MSSM expressions in the limit $\cos\theta_A \to 1$, $\cos\theta_{13} \to 1$, and $\cos\theta_{23} \to 1$. We therefore display these mixing angles in Figure~\ref{fig:CosThetaA} for the NMSSM scans as a function of the input parameter $m_{H^\pm}$. The colour coding corresponds to exclusion by different constraints, which are applied in the order indicated by the legend. This means that the allowed points are displayed on top. Given the sensitivity to the SUSY scenario of   $\rm{BR}(B\to X_s\gamma)$ and $\rm{BR}(B_s\to \mu^+\mu^-)$ these constraints are applied first.
From the figure, we note that for small $m_{H^\pm}$ the points in parameter space that pass the collider and flavour constraints have small mixing with the singlet states, meaning that the couplings are MSSM-like, whereas for larger $m_{H^\pm}$ the mixing with the singlet states, $\sin\theta_A$  and $\sin\theta_{23}$, can be both large and small. The same trends are visible in both scenarios, although this is more clear in scenario (A) since more points pass the constraints in this case.

\subsection{Production of neutral Higgs bosons}
The cross sections for the dominant production modes of the neutral NMSSM Higgs bosons can be obtained by combining the reduced couplings given in Table~\ref{tab:couplings} with the production cross sections for SM Higgs bosons. In addition to the reduced couplings defined in  Table~\ref{tab:couplings} we also need the reduced $\Phi gg$ couplings. These are defined such that the NMSSM cross sections for $\Phi$ production can be obtained from the SM cross sections by multiplying with the reduced coupling squared, thus including the contributions of sparticles appearing in the loop.

\begin{figure}
\centering
\includegraphics[width=0.5\columnwidth]{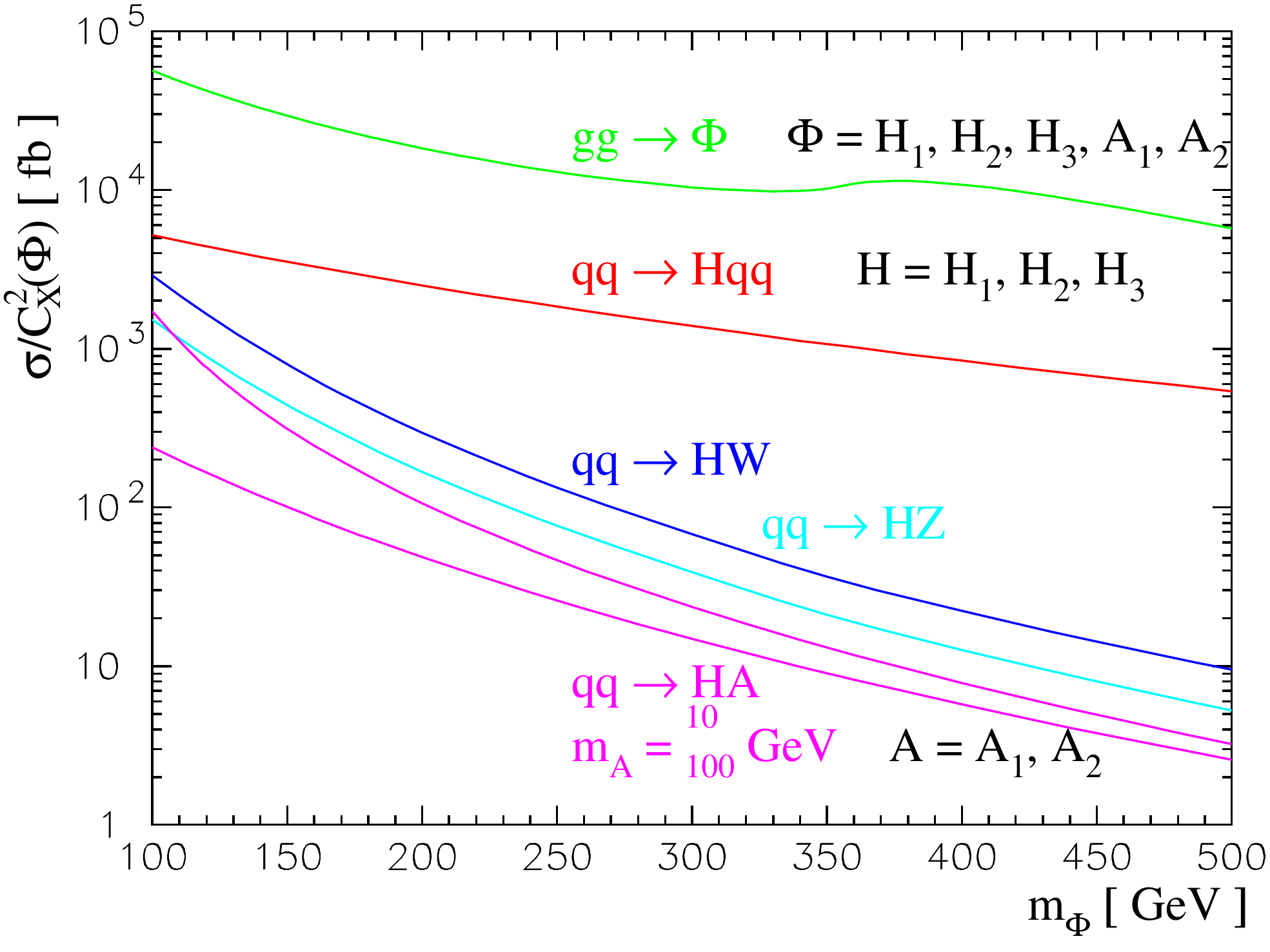}
\caption{The cross section for various Higgs production processes at the LHC ($\sqrt{s}=14\TeV$). The corresponding NMSSM cross sections can be obtained by multiplying with the reduced couplings squared.}
\label{fig:sigma_h_SM}
\end{figure}

For completeness we give some of the most relevant SM cross sections\footnote{The SM cross sections in  Figure~\ref{fig:sigma_h_SM} have been obtained from the compilation for the TeV4LHC workshop which can be found at \url{http://maltoni.web.cern.ch/maltoni/TeV4LHC/SM.html}. In short the $gg \to H$ cross section is based on the NNLO calculation in the large top-mass limit~\cite{Harlander:2002wh,Anastasiou:2002yz} including soft-gluon resummation to NNLL~\cite{Catani:2003zt}. The vector boson fusion cross section has been calculated to NLO~\cite{Han:1992hr,Berger:2004pca,Figy:2003nv} using the {\tt MCFM} program \url{http://mcfm.fnal.gov} and the Higgs-strahlung cross section is calculated to NNLO in QCD~\cite{Brein:2003wg} and NLO in EW~\cite{Ciccolini:2003jy}. For more details on the calculations we refer to the website given above.} in Figure~\ref{fig:sigma_h_SM} together with the leading order cross section\footnote{This cross section has been obtained using {\tt MadGraph/MadEvent}~\cite{Alwall:2007st} with the {\tt CTEQ6L1} parton densities~\cite{Pumplin:2002vw} and using $\sqrt{s}$ as factorisation scale.} for the associated $A_i H_j$ production process $pp \to Z^* \to A_i H_j$ (the latter is given for the two cases $m_{A_i}=10$ and $m_{A_i}=100$ GeV).

In addition to the modes displayed in the figure the neutral Higgs bosons can also be produced together with heavy quarks through processes such as $b\bar{b} \to \Phi$ and $gg \to t\bar{t}\Phi$ with SM cross sections of similar magnitude as the Higgs-strahlung ones ($q\bar{q} \to H V$). In principle these cross sections can be enhanced with a factor $\sim \tan\beta$ compared to the SM. However, in the following we will concentrate on the dominant production processes and those which are of special interest for a light $A_1$ and/or $H^\pm$ and at the same time give charged leptons for signal tagging. 

\begin{figure}
\centering
\includegraphics[width=0.48\columnwidth]{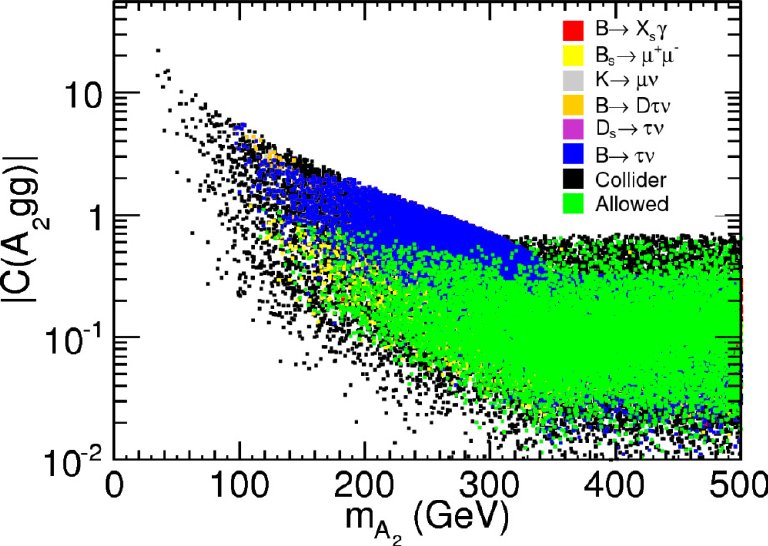}\;\;\;
\includegraphics[width=0.48\columnwidth]{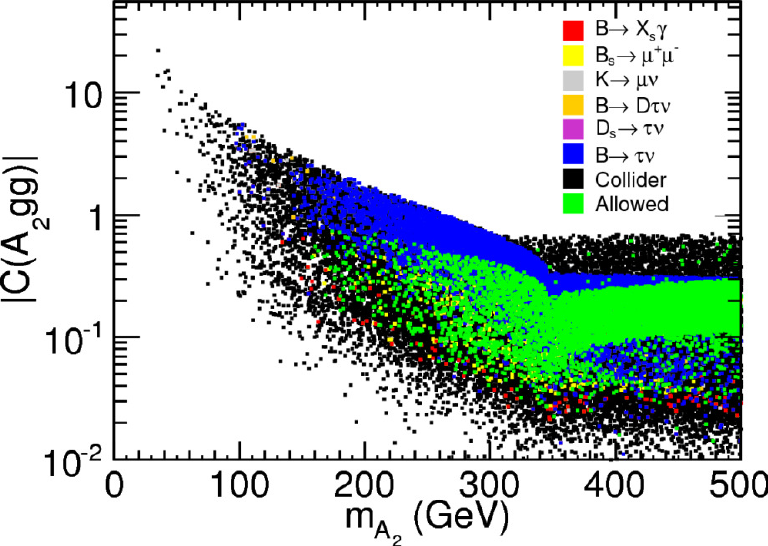}
\includegraphics[width=0.48\columnwidth]{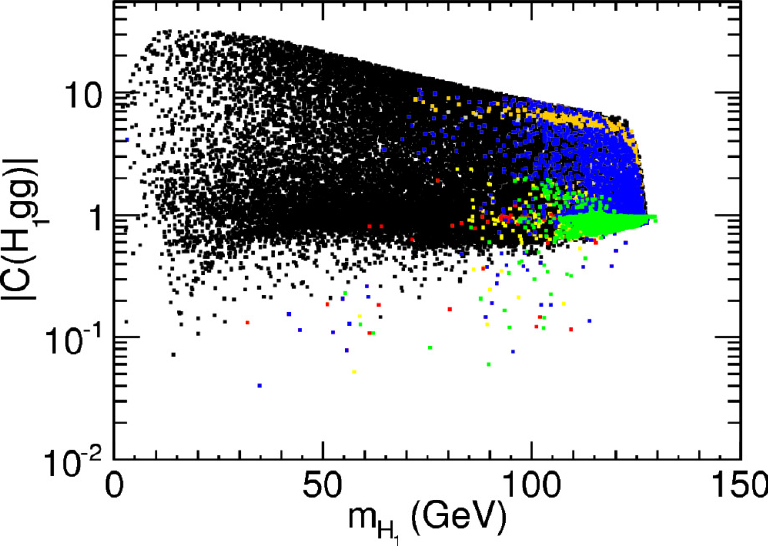}\;\;\;
\includegraphics[width=0.48\columnwidth]{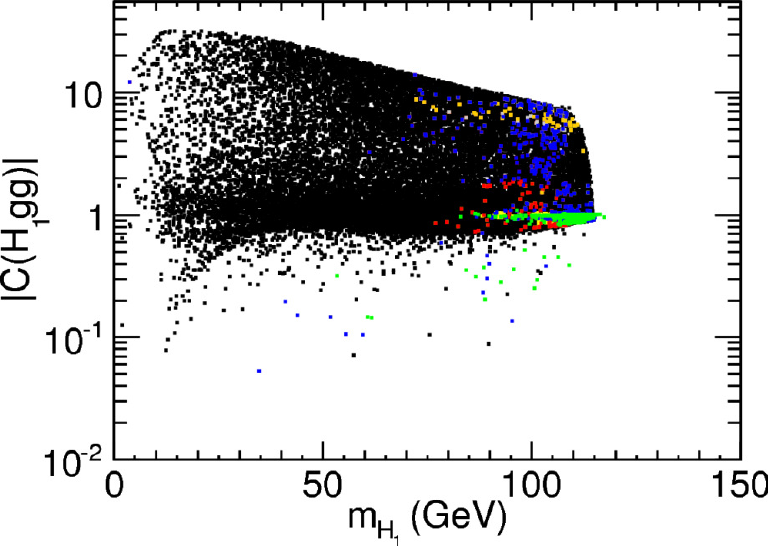}
\includegraphics[width=0.48\columnwidth]{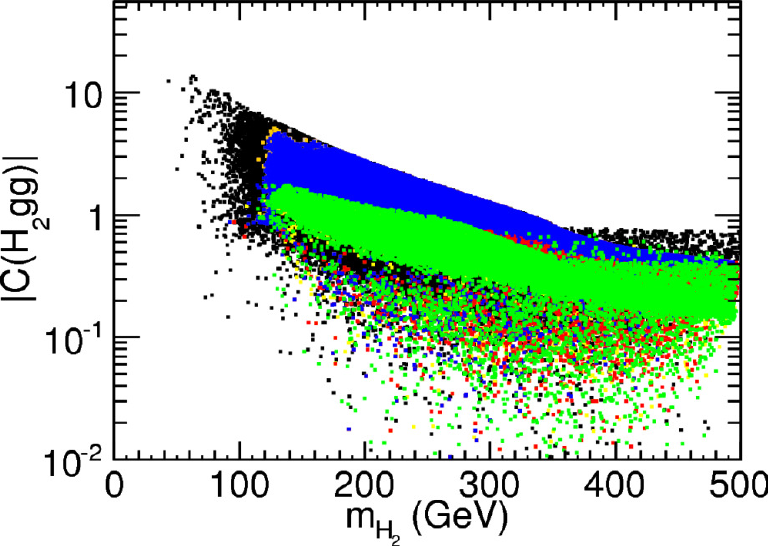}\;\;\;
\includegraphics[width=0.48\columnwidth]{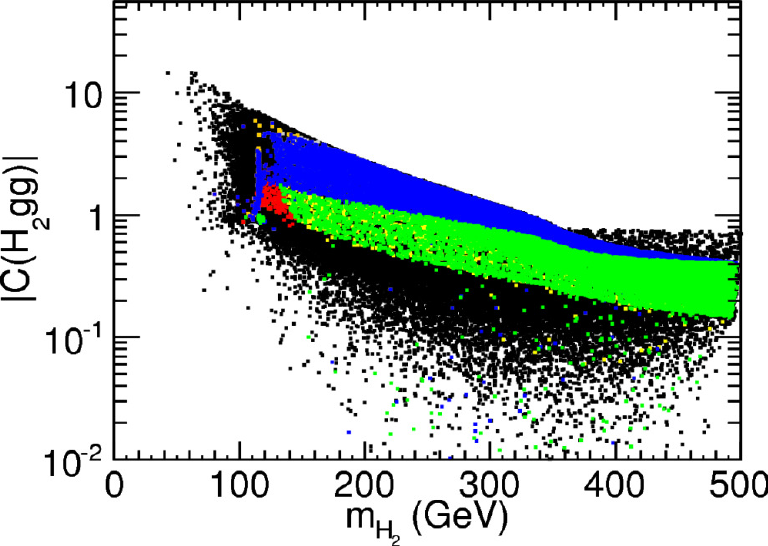}
\caption{The reduced couplings $A_2 gg$ (upper), $H_1 gg$ (middle), and $H_2 gg$ (lower) in the NMSSM relative to the SM for scenario (A) (left) and scenario (B) (right). The constraints are applied in the order indicated by the legend.}
\label{fig:phigg}
\end{figure}
As is clear from Figure~\ref{fig:sigma_h_SM} the $gg$ production mode dominates for SM Higgs masses above $100$ GeV. In addition, since the reduced couplings $H_i VV$ are always smaller than one, the same will also be true in the NMSSM as long as the reduced $\Phi gg$ (with $\Phi =H_i,A_i$) couplings are not too small. The reduced couplings of the relevant Higgs states are shown in Figure~\ref{fig:phigg}. We leave out $H_3$ as mentioned above, and also $A_1$ since it is difficult to use single $A_1$ production experimentally when $A_1$ is light.\footnote{There could be some hope to extract a signal in $gg\to  A_1b\bar{b}$ using the $b$ jets to tag the event \cite{Almarashi:2010jm}.} Since the reduced $\Phi gg$ couplings are dominated by the $\Phi b\bar{b}$ couplings for high $\tan\beta$, the reduced couplings can in principle be of this order. However, after applying the flavour constraints (chiefly $B\to\tau\nu_\tau$ and $B\to D\tau\nu_\tau$), these couplings are close to one in both scenarios for $H_1$ and $H_2$, whereas for $A_2$ they are typically smaller than one due to the extra suppression by $\sin\theta_A$. 

\begin{figure}
\centering
\includegraphics[width=0.48\columnwidth]{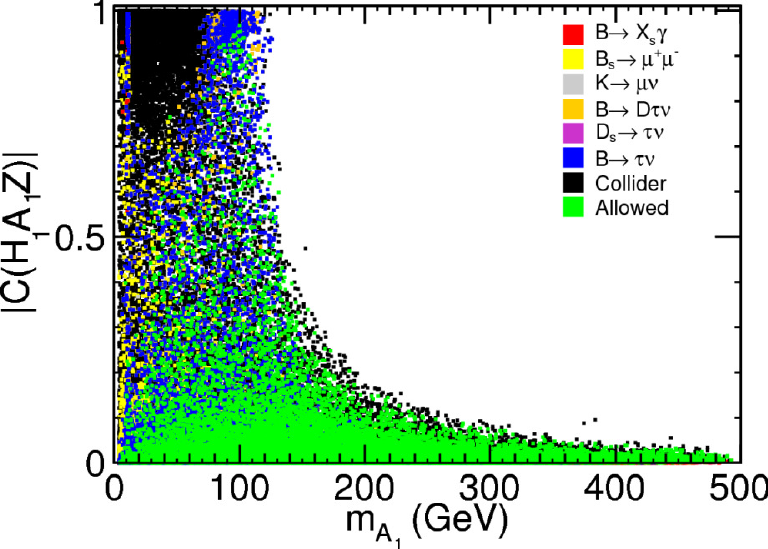}\;\;\;
\includegraphics[width=0.48\columnwidth]{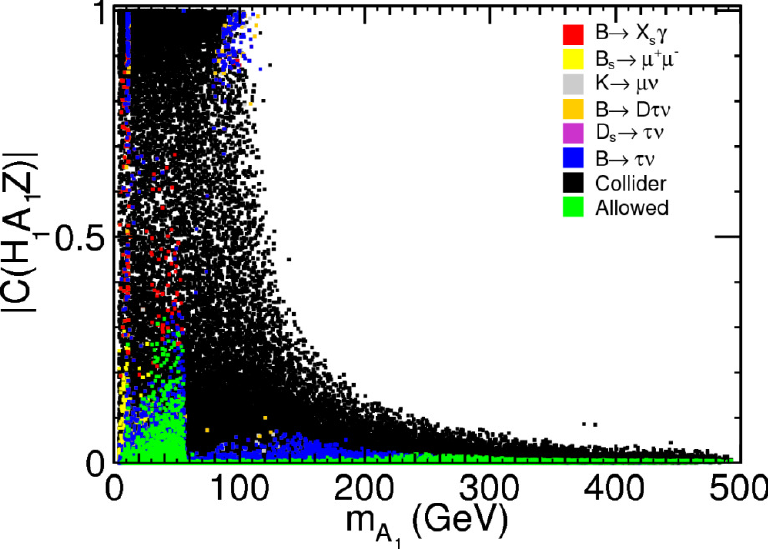}
\includegraphics[width=0.48\columnwidth]{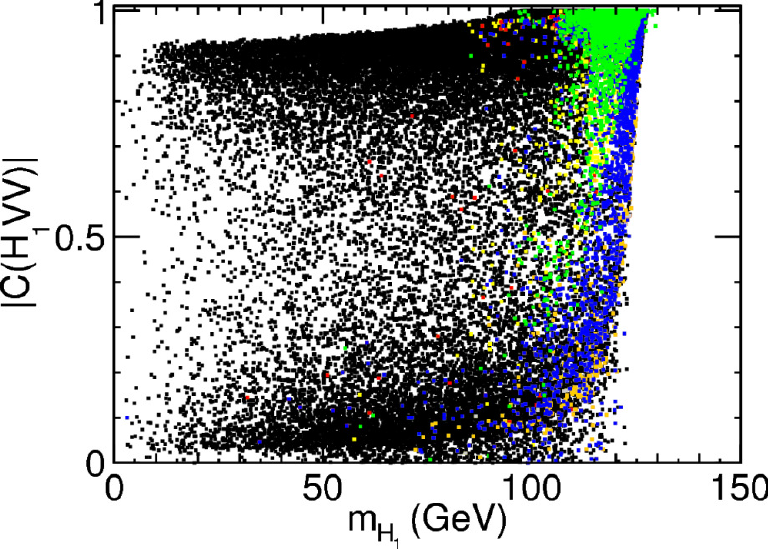}\;\;\;
\includegraphics[width=0.48\columnwidth]{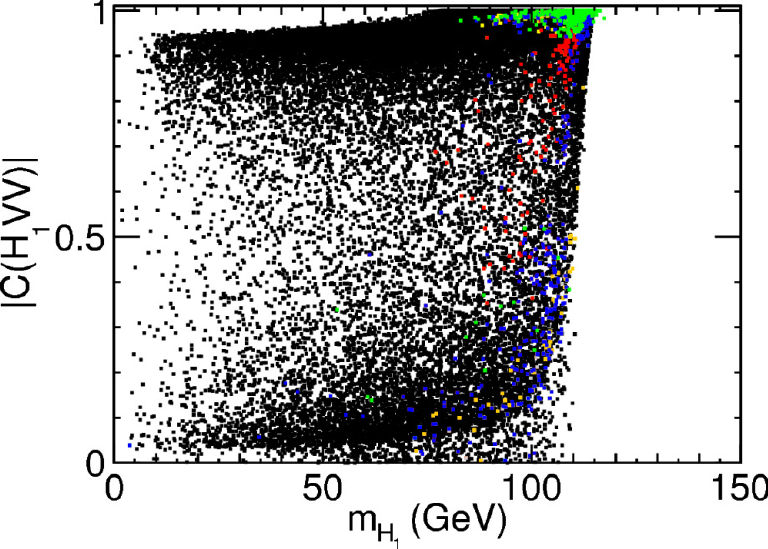}
\includegraphics[width=0.48\columnwidth]{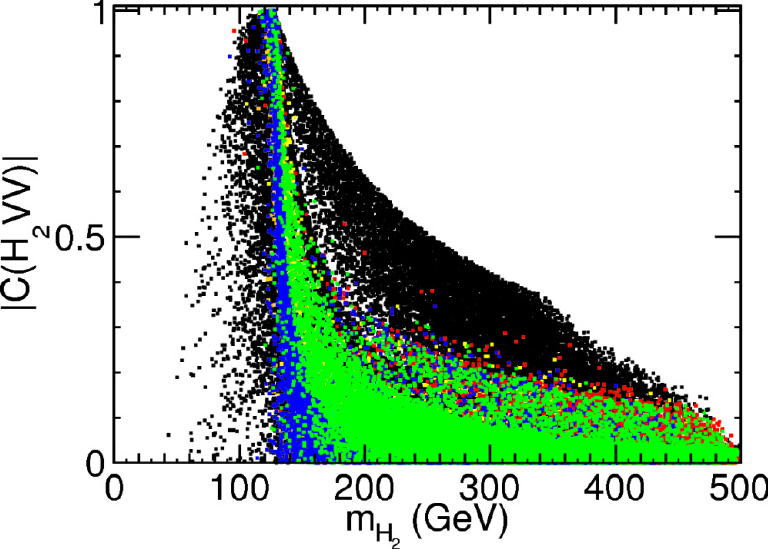}\;\;\;
\includegraphics[width=0.48\columnwidth]{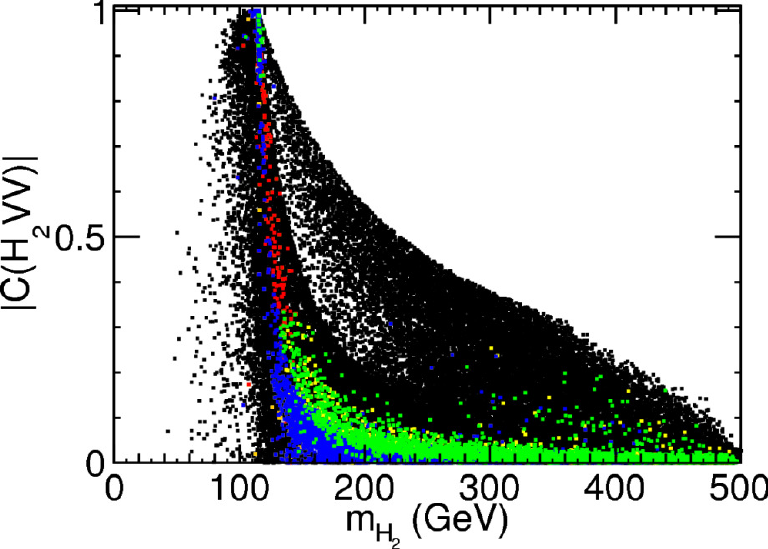}
\caption{The reduced couplings $A_1H_1Z$ (upper), $H_1VV$ (middle), and $H_2VV$ (lower) in the NMSSM relative to the SM for scenario (A) (left) and scenario (B) (right). The constraints are applied in the order indicated by the legend.}
\label{fig:phiVV}
\end{figure}
As has been already mentioned, the cross sections for production of neutral Higgs bosons through processes with electroweak bosons are typically much smaller than the ones through gluon fusion. Since both $C(H_1VV)$ ands $C(H_1gg)$ are close to one, both production modes are SM-like, 
so the $H_1 \to A_1A_1$ decays may degrade the possibilities of discovery in the standard mode which is already very difficult in the relevant $m_{H_1}$ range. In addition, using the $gg \to H_1 \to A_1A_1$ channel is presumably difficult experimentally, whereas vector boson fusion is more hopeful \cite{Belyaev:2008gj,Rottlander:2008zz,*Rottlander:2009zz}. The electroweak production may be of phenomenological interest since it provides additional signatures for tagging (backward/forward jets in the case of vector bosons fusion and additional electroweak gauge bosons in the case of Higgs-strahlung).

Vector boson fusion and Higgs-strahlung production of $H_2$ are only relevant for $m_{H_2} \lesssim 150$~GeV in scenario (A) since $C(H_2VV)$ is too small for large masses after applying the collider constraints. In scenario (B) it is almost never relevant, although this depends mainly on the exclusion by $b \to s \gamma$. This also means that the decays $H_2 \to A_1 Z$ are never relevant for this production mode (see Figure \ref{fig:brHHV}).
 
In the case of a light $A_1$, the cross section for associated $A_1H_1$ production may be comparable to production with a vector boson, in particular for small $H_1$ masses as shown in Figure~\ref{fig:sigma_h_SM}. 
The relevant reduced couplings for these cases are displayed in Figure~\ref{fig:phiVV}. 
As can be seen in the figure the $H_1VV$ coupling is typically large in both scenarios after applying the flavour constraints. The opposite is true for the $H_2VV$ and $A_1H_1Z$ couplings in scenario (A), except for small $H_2$ and $A_1$ masses respectively where again the reduced couplings can be large. 

\begin{figure}
\centering
\includegraphics[width=0.4\columnwidth]{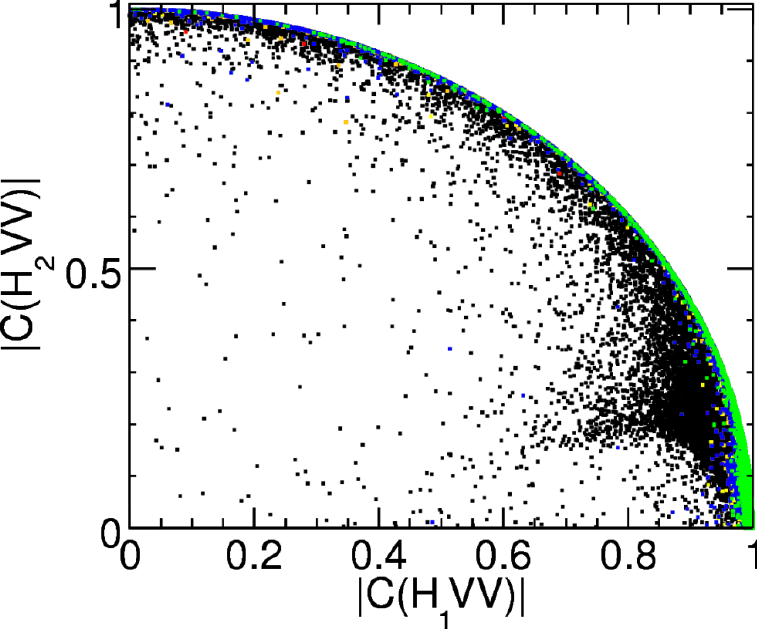}\;\;\;\;\;
\includegraphics[width=0.4\columnwidth]{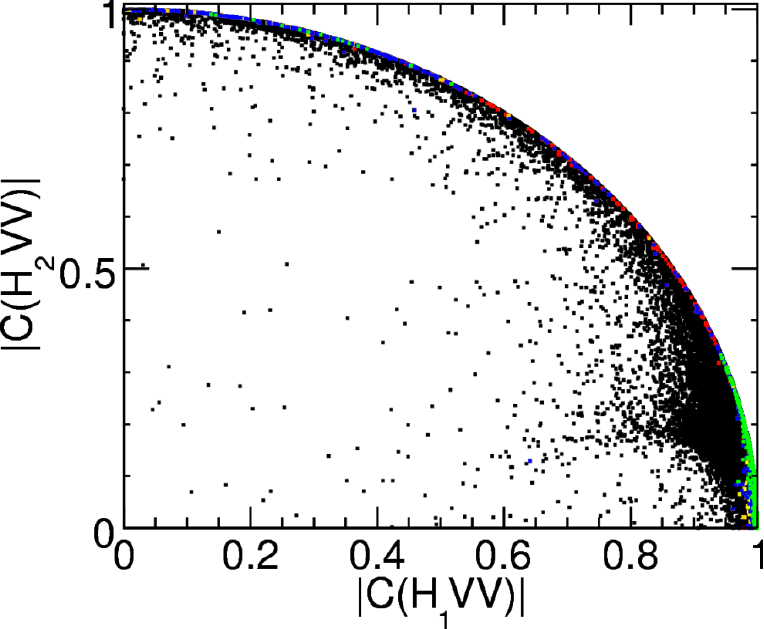}
\caption{The relation between the $H_1VV$ and $H_2VV$ couplings in the NMSSM. Scenario (A) left, scenario (B) right.}
\label{fig:phiVV_ratio}
\end{figure}

We would like to emphasise that even though the $H_1VV$ and $H_2VV$ couplings in principle are smaller in the NMSSM than in the MSSM due to mixing with the singlet state, the sum rule $\sum C(H_i VV)^2 = 1$ is often saturated by the two lightest states (similarly to the MSSM). This holds especially when flavour and collider constraints are applied, as illustrated in  Figure~\ref{fig:phiVV_ratio}.

\subsection{Decays of neutral Higgs bosons}
We turn now to the decays of the neutral Higgs bosons. Again we concentrate on those modes which are dominant or of special interest when $A_1$ and/or $H^\pm$ are light, still keeping in mind the need for charged leptons for tagging.
\begin{figure}
\centering
\includegraphics[width=0.48\columnwidth]{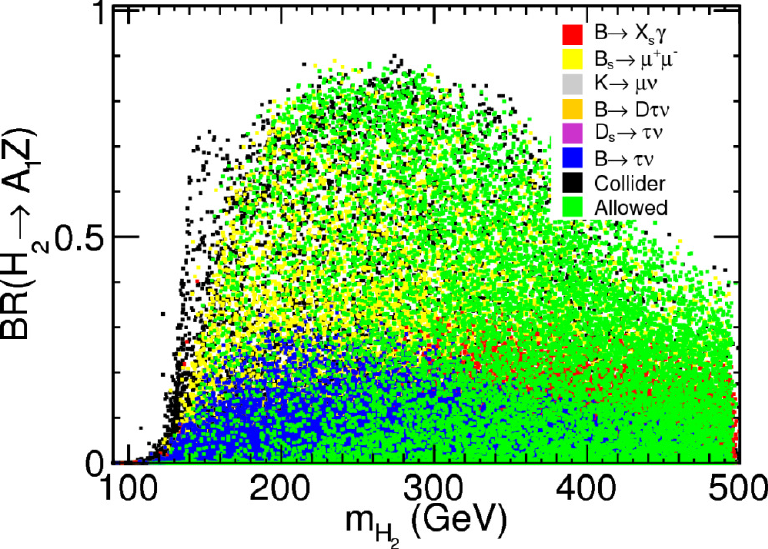}\;\;\;
\includegraphics[width=0.48\columnwidth]{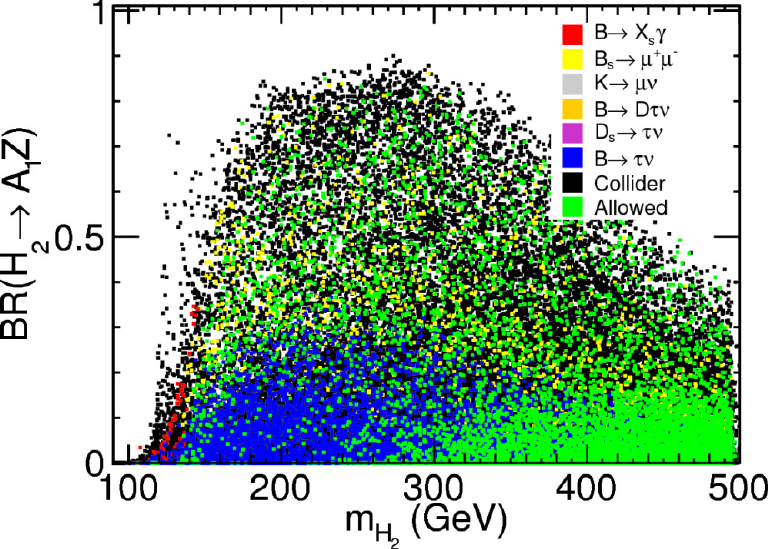}
\includegraphics[width=0.48\columnwidth]{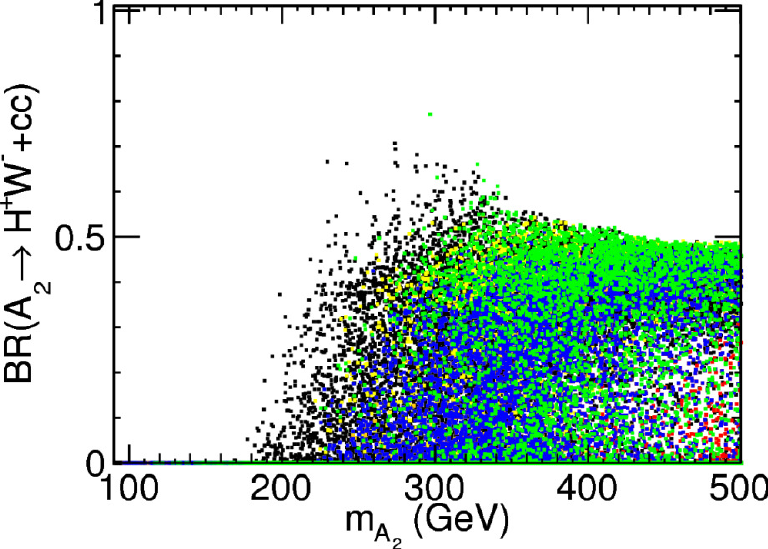}\;\;\;
\includegraphics[width=0.48\columnwidth]{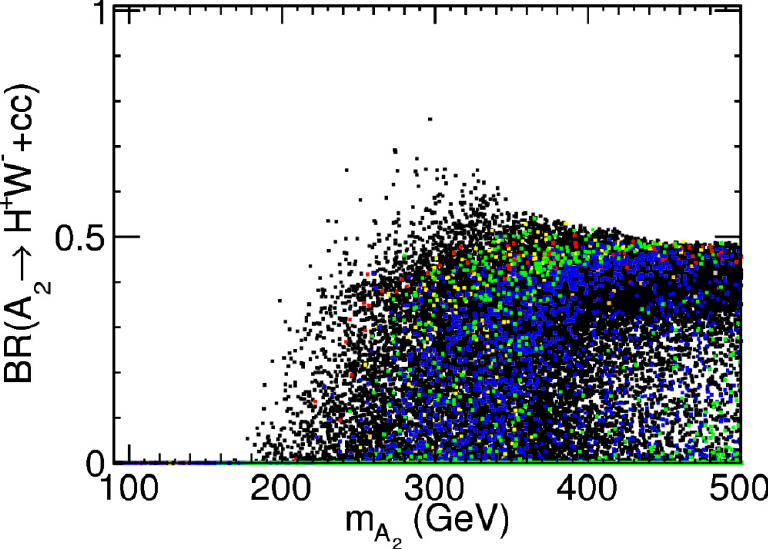}
\includegraphics[width=0.48\columnwidth]{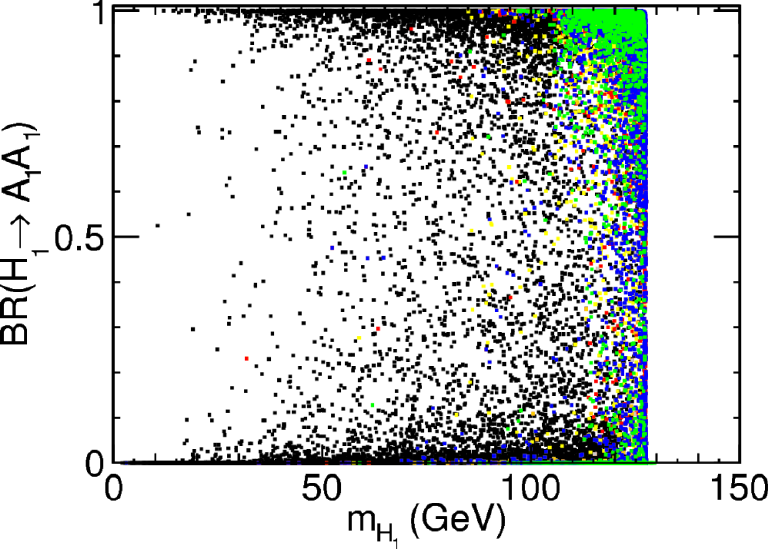}\;\;\;
\includegraphics[width=0.48\columnwidth]{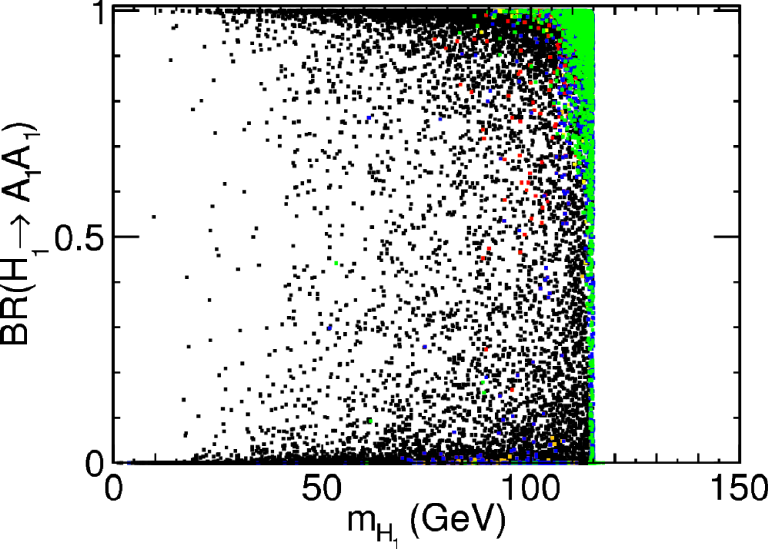}
\caption{The branching ratios for $H_2 \to A_1 Z$ (upper), $A_2 \to H^\pm W$ (middle), and   $H_1 \to A_1 A_1$ (lower) in the NMSSM. Scenario (A) left, scenario (B) right.}
\label{fig:brHHV}
\end{figure}
In the case of Higgs boson production through gluon fusion the decay modes of most interest to us are $H_2 \to A_1 Z$ (the branching ratio for $H_1 \to A_1Z$ is small, always at the level $10^{-2}$ or less) and possibly $ H_1  \to  A_1 A_1 $, as well as
$A_2 \to H^\pm W\to A_1 W^+ W^-$ or possibly $A_2 \to H_1 Z\to  A_1 A_1 Z$. The same decay modes are also of interest for the production through electroweak gauge bosons. We note in passing that similarly to the MSSM, $\rm{BR}(A_1 \to \tau^+ \tau^-)\simeq 0.1$ above $b\bar{b}$ threshold with $A_1 \to b\bar{b}$ saturating the width. Below the $b\bar{b}$ threshold $\rm{BR}(A_1 \to \tau^+ \tau^-)\simeq 0.9$. 

Figure~\ref{fig:brHHV} shows that the branching ratio $H_2 \to A_1 Z$ can be dominant. Together with the large cross section for $H_2$ production this makes an interesting channel for the LHC. The same is true for the decay $A_2 \to H^\pm W$, whereas the branching ratio for  $A_2 \to H_1 Z$ turns out to be much smaller. This is mainly due to the comparatively smaller number of final state particles (not shown). Finally, we see that the  $H_1 \to A_1 A_1$ decay can be very dominant as soon as the masses are such that the decay is open. These conclusions depend only weakly on the SUSY scenario, as is clear from the small differences observed in Figure~\ref{fig:brHHV}. Combining production and decay, the product $C(A_2gg)^2\times \mathrm{BR}(A_2\to H_1Z)$ is in the best case of the order $10^{-2}$. The same is true for $C(A_2gg)^2\times \rm{BR}(A_2\to H^\pm W^\mp)$, which could even so be relevant for charged Higgs production since this mode is of a similar magnitude as the other $H^\pm$ production modes (to be discussed next). 

\begin{figure}
\centering
\includegraphics[width=0.5\columnwidth]{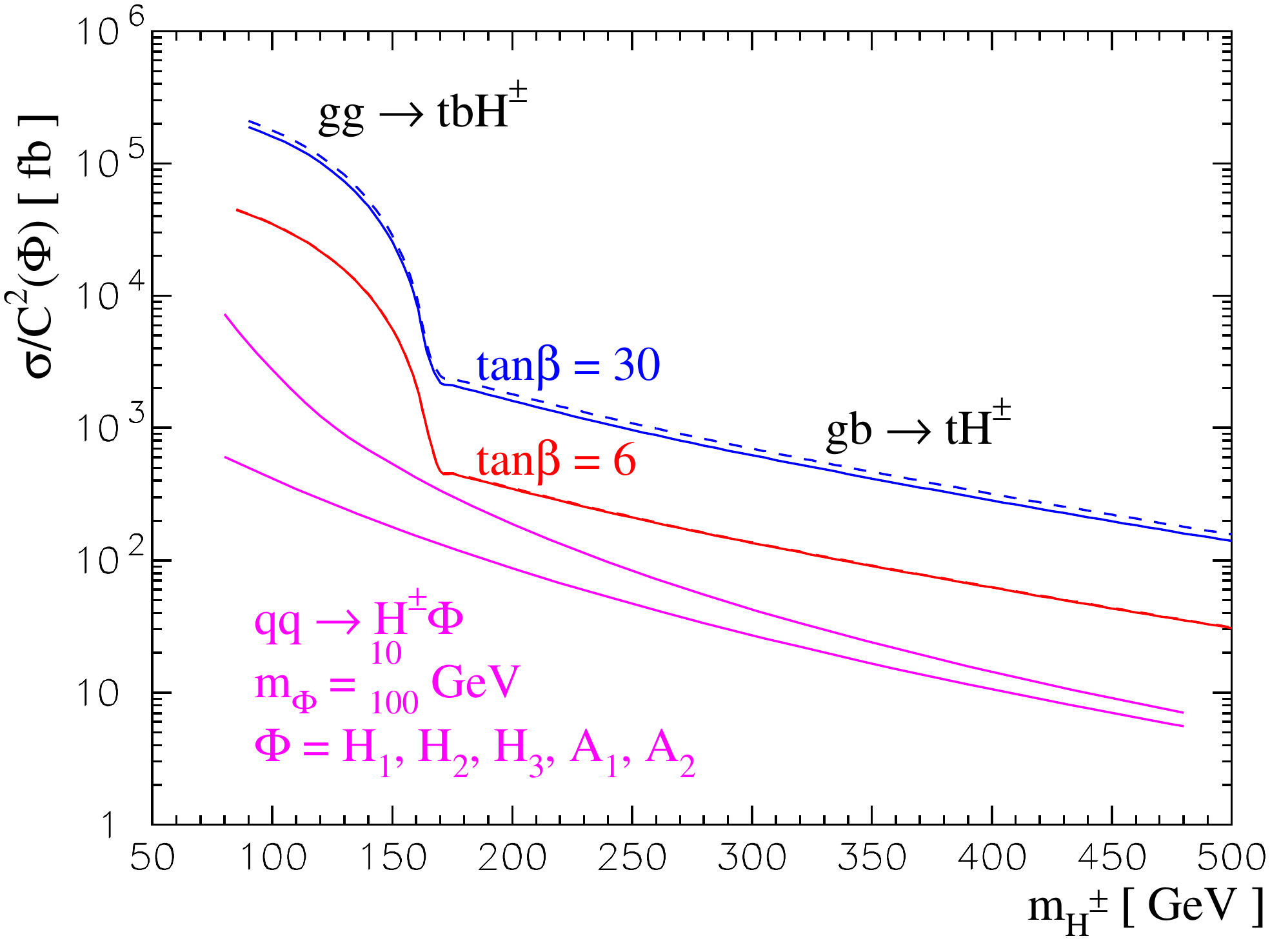}
\caption{The cross sections for various charged Higgs production processes at the LHC ($\sqrt{s}=14\TeV$). The values for associated production $H^\pm\Phi$ should be multiplied with the reduced couplings squared to obtain the NMSSM results. The solid line is for scenario (A) and the dashed one for scenario (B).}
\label{fig:sigma_HPM}
\end{figure}

\subsection{Production and decay of charged Higgs bosons}
The main production mechanism for heavy charged Higgs bosons ($m_{H^\pm} > m_t$) at the LHC is through the twin processes $gb\to tH^\pm$ and $gg \to tbH^\pm$, depending on whether one is using a four or a five flavour scheme for the parton densities. In addition the $gg \to tbH^\pm$ process also gives the dominant production mechanism for light charged Higgs bosons ($m_{H^\pm} < m_t$) which proceeds through top-pair production with one of the tops decaying to a charged Higgs. In both the NMSSM and the MSSM the magnitude of the cross section is determined by the $H^\pm tb$  coupling, $g_{H^\pm tb} \propto (m_t\frac{1-\gamma_5}{2}\cot\beta+m_b\frac{1+\gamma_5}{2}\tan\beta)$, giving a minimum around $\tan\beta=\sqrt{m_t/m_b}\sim 7$. The production cross section above the $t\bar{t}$ threshold has been calculated to NLO both in the five~\cite{Zhu:2001nt,*Plehn:2002vy,*Berger:2003sm} and four~\cite{Dittmaier:2009np} flavour schemes. In Figure~\ref{fig:sigma_HPM} we show the cross section calculated in the five flavour scheme as given by the parametrisation contained in {\tt FeynHiggs}~\cite{Frank:2006yh,*Degrassi:2002fi,*Heinemeyer:1998np,*Heinemeyer:1998yj}. Below the $t\bar{t}$ threshold we have used the $t\to bH^+$ branching ratios calculated with the same program, based on the NLO calculation~\cite{Korner:2002fx,*Carena:1999py}, and multiplied with a $t\bar{t}$ production cross section of $900$~pb~\cite{Cacciari:2008zb}.

\begin{figure}
\centering
\includegraphics[width=0.48\columnwidth]{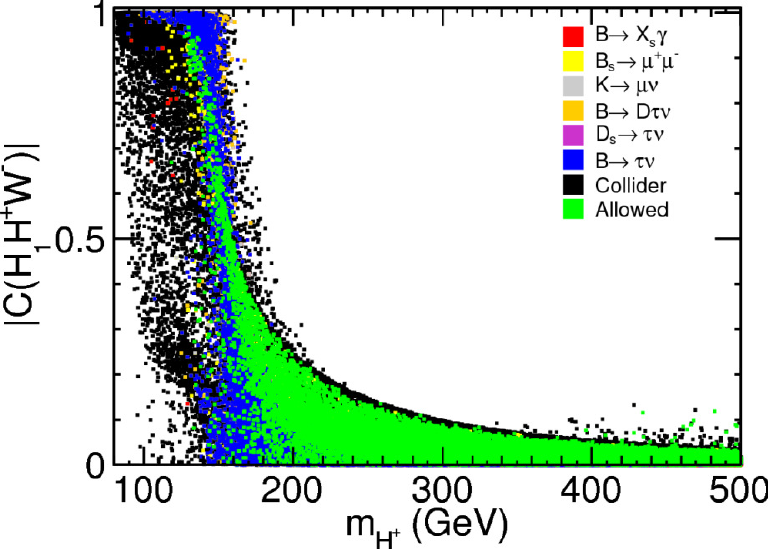}
\includegraphics[width=0.48\columnwidth]{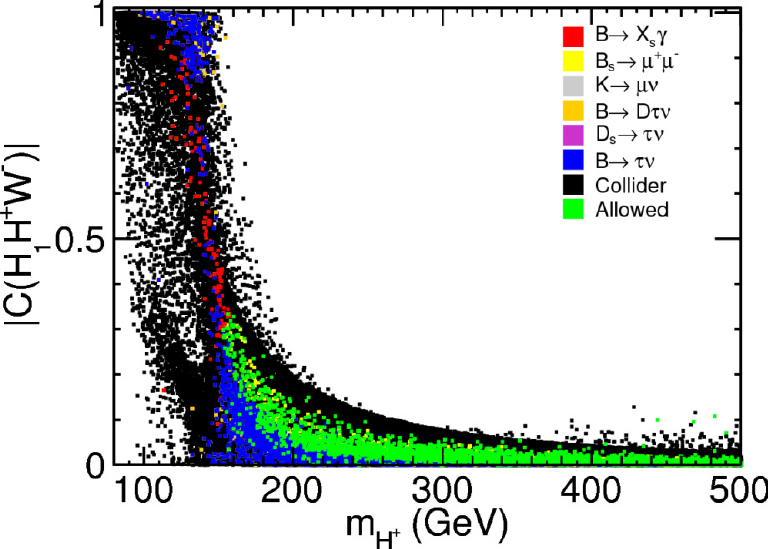}\\
\caption{The reduced $H_1H^+W^-$ coupling in the NMSSM for scenario (A) (left) and scenario (B) (right).}
\label{fig:PhiHpW}
\end{figure}

In addition to this standard production mechanism, the possibility of a light $A_1$ (and possibly a light $H_1$) in the NMSSM means that associated $H^\pm\Phi$ (with $\Phi=A_1,H_1$) production can also be large enough to be of phenomenological interest. This is explored in some detail in \cite{Akeroyd:2007yj} (without the explicit application of flavour constraints). Figure~\ref{fig:sigma_HPM} gives the LO cross section\footnote{Similarly to the associated $HA$ production the $H^+\Phi$ cross section has been obtained using {\tt MadGraph/MadEvent}~\cite{Alwall:2007st} with the {\tt CTEQ6L1} parton densities~\cite{Pumplin:2002vw} and $\sqrt{s}$ as factorisation scale. For an NLO calculation and analysis of this process in the MSSM, see \cite{Cao:2003tr}.} apart from the reduced couplings $H_1H^+W^-$ and $A_1H^+W^-$, which are given by Figures~\ref{fig:PhiHpW} and \ref{fig:CosThetaA}, respectively (note that the reduced $A_1H^+W^-$ coupling is equal to $\cos\theta_A$). From this it is clear that in scenario (A) both production modes can be of interest, at least for light charged Higgs bosons, whereas in scenario (B) the cross section for $pp \to H_1H^\pm$ is probably too small to be of phenomenological interest. The dampening of $C(H_1H^+W^-)$ for large $m_{H^\pm}$ is a consequence of the sum rule $C(H_iVV)^2 + C(H_iH^+W^-)^2 + S_{i3}^2 =~1$ \cite{Akeroyd:2007yj} (cf.~Figure \ref{fig:phiVV}). 

\begin{figure}[t!]
\centering
\includegraphics[width=0.48\columnwidth]{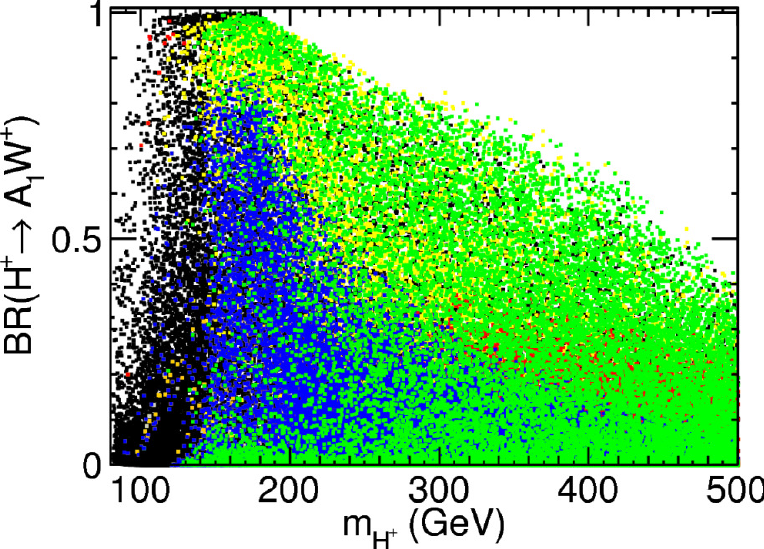}
\includegraphics[width=0.48\columnwidth]{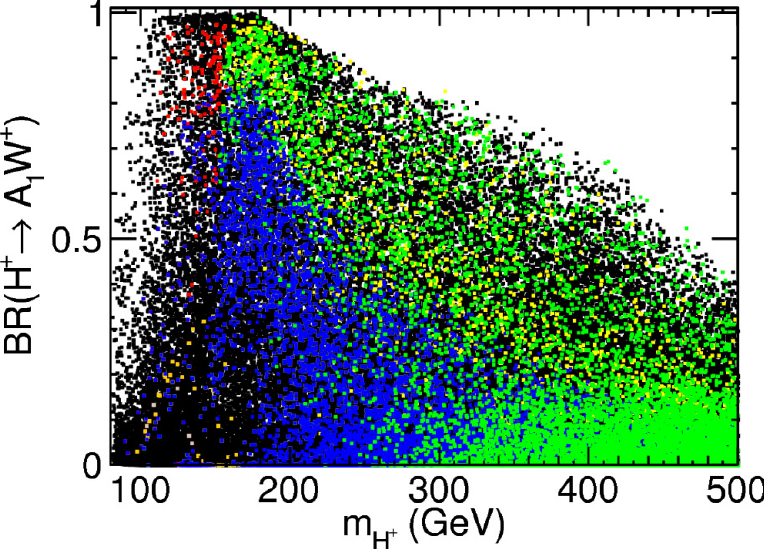}
\caption{The branching ratio for $H^+ \to A_1W^+$ in the NMSSM (computed for off-shell $W^+$) in scenario (A) (left) and scenario (B) (right).}
\label{fig:brhp_a1w_of_mHp}
\end{figure}

Finally the presence of a light  $A_1$ also opens up the possibility of a substantial branching ratio for $H^+ \to A_1W^+$ decays. As is shown in 
Figure~\ref{fig:brhp_a1w_of_mHp}, this remains the case in both scenarios also after the flavour constraints have been applied.

\section{Summary and conclusions}
\label{sect:Summary}

To summarize, we have considered NMSSM extensions of well-known (and widely used) MSSM benchmark scenarios for Higgs physics. This provides a `fair' method to compare the two models without resorting to theoretical bias (such as fine-tuning measures). Our numerical results are presented for two specific benchmark scenarios, but the method and conclusions are easily generalized to other scenarios.
Using data from LEP, the Tevatron, and flavour physics experiments, we determine the allowed regions of parameter space for the Higgs bosons masses. In both benchmark scenarios we find that the collider limits allow for one or more Higgs bosons which are substantially lighter in the NMSSM ($m_{H_1}<m_Z$, $m_{A_1}\lesssim 10\GeV$, $m_{H^\pm}\sim 90\GeV$) compared to the limits that apply in the MSSM ($m_{h}>m_Z$, $m_{A}>m_Z$, $m_{H^\pm}\gtrsim 120\GeV$). We also find that the low $\tan\beta$ region excluded in the MSSM is reopened in the NMSSM. These conclusions hold also when including present limits from channels geared specifically towards detection of light NMSSM Higgs bosons, such as $H_1\to A_1A_1\to 4\tau$ and $H^\pm\to W^\pm A_1$. 

Direct limits from searches for charged Higgs bosons could in principle be more general than those from neutral Higgs searches, since no new charged degrees of freedom are introduced in the NMSSM. Unfortunately the current sensitivity is not at the level necessary to provide any significant exclusion beyond the LEP results of $m_{H^\pm}\gtrsim 80\GeV$, except for very high (and very low) $\tan\beta$. Improved exclusion limits on $H^\pm$ would be useful to constrain the NMSSM parameter space further. In fact, not even the present MSSM limits on $H^\pm$ are completely unavoidable, since the additional decay channel $H^\pm\to W^\pm A_1$ could degrade the performance of the standard searches. We therefore strongly recommend that this channel is considered for LHC analyses in order to fill this gap. The preliminary CDF results on this mode should serve as encouragement and a challenge to the LHC collaborations.

Flavour physics experiments are complementary to those at high-energy colliders, since the observables measured there can be exploited to constrain the Higgs sector indirectly far beyond the energy scale which is directly accessible. With constraints from flavour physics included, we still find that lighter charged Higgs bosons than in the MSSM can be accommodated in the NMSSM. However, this typically requires that the underlying SUSY scenario is tuned to cancel the $H^\pm$ contributions to $B\to X_s\gamma$ transitions. Another observable of importance in the NMSSM is the $B_s\to \mu^+\mu^-$ decay which is sensitive to the presence of a light $A_1$. This leads to a larger variation between the new physics contributions in the MSSM and the NMSSM, since a light $A_1$ can be realized independently of $m_{H^\pm}$. In the NMSSM, we find that the constraints from $B_s\to \mu^+\mu^-$ can  be excluding down to moderate $\tan\beta$ values. It would therefore be useful to extend the calculation beyond the usual high $\tan\beta$ limit to study this region in more detail. An experimental measurement of $B_s\to \mu^+\mu^-$ would of course also be very welcome to further constrain (or confirm) physics beyond the SM. For $\tan\beta\gtrsim 10$, both $B_s\to \mu^+\mu^-$ and the leptonic decays of pseudoscalar mesons (e.g.~$B_u\to \tau\nu$) rule out the light $H^\pm$ scenario. Since the latter are sensitive to tree-level $H^\pm$ exchange, the derived limits on ($m_{H^\pm}, \tan\beta$) remain robust in the NMSSM (as they are also with respect to changes of MSSM scenario).
\newpage
Having applied the collider and flavour constraints, we evaluated quantities of interest for LHC Higgs phenomenology. Looking at the results for the reduced couplings relevant to the main production modes, we find that the CP-even Higgs sector in general is similar to that in the MSSM. This means that gluon fusion remains the dominant mode of production and that the coupling of the heaviest CP-even Higgs boson to vector bosons is very small. The main phenomenological difference here is therefore the additional decay $H_1\to A_1A_1$, which can be large in both scenarios also after constraints are applied. In the CP-odd sector, the production $gg\to A_2$ is suppressed due to singlet mixing. Nevertheless we find that the mode $gg\to A_2\to H^\pm W^\mp$ deserves further investigation as a channel not present in the MSSM. For the direct production of charged Higgs bosons, an interesting alternative to the standard mode is offered by $pp\to H^\pm A_1$, in particular when a light $A_1$ is mostly doublet which gives a reduced coupling of $\mathcal{O}(1)$. The same coupling also governs the decay $H^\pm\to  W^\pm A_1$ which is often dominant when kinematically allowed.

In conclusion the LHC could be facing scenarios which differ substantially from the MSSM in the Higgs sector, while the sparticle sector is effectively unchanged. It is important to be aware of this possibility when designing Higgs search strategies; either trying to maintain model independence or by including modes which are not relevant in the MSSM. This is of course primarily motivated by the fact that Higgs discovery can not be guaranteed in the NMSSM (unlike the MSSM). Taking a more optimistic perspective, we would like to stress the importance that no channel is left behind since the NMSSM can offer a much richer accessible Higgs phenomenology that deserves to be fully exploited. 

\section*{Acknowledgements}
We thank Andrew Ivanov and William Johnson of the CDF experiment for providing us with their results on the $H^+\to W^+A$ channel. JR acknowledges partial support from the Swedish Research Council under contract 621-2008-4219.

\bibliography{paper2}

\end{document}